\documentclass[11pt]{article}

\usepackage[]{acl}

\usepackage{times}
\usepackage{latexsym}

\usepackage[T1]{fontenc}
\usepackage[utf8]{inputenc}
\usepackage{microtype}
\usepackage{inconsolata}
\usepackage{graphicx}
\usepackage{booktabs}

\usepackage{tcolorbox}

\usepackage{multirow}

\usepackage{makecell}

\usepackage{tabularx}
\usepackage{array}
\usepackage{makecell}
\usepackage{stfloats}
\usepackage{float}
\usepackage{placeins}
\usepackage{multirow}

\title{Cleaner Speech, Weaker Generalization: Revisiting Pitt-Derived Benchmarks for Alzheimer’s Disease Detection}

\author{
\textbf{Luqi Sun}$^{1}$,
\textbf{Shreeram Suresh Chandra}$^{1\dagger}$,
\textbf{Lin Zhang}$^{1\dagger}$, 
\textbf{You-Jin Li}$^{2}$,\\
\textbf{Brian~MacWhinney}$^{3}$,
\textbf{Yu Tsao}$^{2}$,
\textbf{Emily Mower Provost}$^{4}$,
\textbf{Berrak Sisman}$^{1}$ \\
$^{1}$ Center for Language and Speech Processing (CLSP), Johns Hopkins University\\
$^{2}$ Research Center for Information Technology Innovation, Academia Sinica\\
$^{3}$ Carnegie Mellon University, $^{4}$ University of Michigan, Ann Arbor \\
\texttt{\{lsun59, sisman\}@jhu.edu} 
}

\begin{document}
\maketitle

\begingroup
\renewcommand\thefootnote{}
\footnotetext{$^{\dagger}$ These two authors contributed equally as second authors.}

\footnotetext{\label{fn:code}We publish our code here: \url{https://github.com/Sleepwalker554/Back_to_Origin}}

\endgroup

\begin{abstract}
Speech-based Alzheimer’s disease (AD) detection increasingly relies on speech-enhanced and curated versions of the Pitt Corpus, where speech enhancement, sample selection, and demographic balancing are often treated as beneficial preprocessing steps. However, whether these transformations improve real-world AD detection or instead affect model generalization and prediction behavior remains unclear. In this work, we revisit the role of speech preprocessing and dataset curation across widely used benchmarks for speech-based AD detection. We evaluate the speech quality of different datasets, the cross-dataset generalization of multiple deep learning models under matched and mismatched enhancement settings, and the behavior of several recent large audio-language models (LALMs). Experimental results show that across multiple supervised speech models, speech-enhanced datasets often improve in-domain performance while reducing robustness in cross-domain evaluation. Matched enhancement between training and test data alleviates, but does not eliminate, this degradation. LALMs show a similar sensitivity: enhanced datasets induce stronger class imbalance and prediction shifts than unprocessed data. These results suggest that speech preprocessing and dataset curation can substantially influence downstream AD detection behavior, indicating that ``cleaner'' speech datasets are not necessarily more reliable for real-world AD detection.

\end{abstract}

\section{Introduction}
Alzheimer’s disease (AD) is the leading cause of dementia worldwide and poses a rapidly growing social and economic burden~\cite{AlzheimerEurope2019, AlzheimersFactsFigures2023}. As the number of individuals living with dementia continues to rise globally~\cite{Scheltens2021}, there is increasing interest in developing robust and cost-effective approaches for early AD detection. Among different modalities, speech has emerged as a promising signal due to its non-invasive nature, low collection cost, and sensitivity to cognitive decline~\cite{literaturereview, literaturereview1}. 

In speech-based AD detection, the Pitt Corpus~\cite{BeckerArchNeurol1994} is the most widely used English benchmark dataset. It contains four tasks, among which the ``Cookie Theft'' task from the Boston Diagnostic Aphasia Examination~\cite{goodglass1983boston} collected spontaneous speech, and therefore has been the most widely used in research. In this task, all participants (both participants with AD and healthy controls) were asked to describe a picture depicting the ``cookie theft'' scene.
Importantly, the Pitt Corpus exists in two versions: the original Pitt-origin recordings and the noise-reduced Pitt version (Pitt), both distributed through DementiaBank~\cite{DementiaBank}. Since the two datasets are otherwise identical, prior studies often do not explicitly distinguish between them, leaving the impact of speech preprocessing on downstream AD detection models largely unexplored.

Beyond Pitt and Pitt-origin, several influential AD detection challenges have introduced additional Pitt-derived datasets, including ADReSS~\cite{ADReSS}, ADReSSo~\cite{ADReSSo}, and ADReSS-M~\cite{LuzMultilingualICASSP2023}. These datasets are all constructed from the ``Cookie Theft'' picture description task in the 
original Pitt-origin dataset, but apply different preprocessing and curation strategies, such as speech enhancement, volume normalization, demographic balancing, and sample selection. While these processing steps are often intended to improve data quality and evaluation fairness, they also alter the original speech distribution and reduce the dataset size.

Despite the widespread use of Pitt-derived datasets such as Pitt, ADReSS, ADReSSo, and ADReSS-M, the effects of speech preprocessing and dataset curation on downstream AD detection remain insufficiently understood. Although these datasets originate from the same ``Cookie Theft'' picture description recordings in Pitt-origin, they apply different preprocessing and selection strategies, including denoising, volume normalization, demographic balancing, and sample filtering. While such processing can improve perceptual speech quality and simplify evaluation, it may also alter the original speech distribution and affect model generalization across datasets and recording conditions. In addition, the reduced dataset size in ADReSS, ADReSSo, and ADReSS-M may increase the risk of overfitting to specific data distributions. As a result, models trained on curated datasets may achieve strong in-domain performance while exhibiting reduced robustness in cross-domain settings.


This paper revisits how speech preprocessing and dataset curation affect AD detection across multiple modeling paradigms. We evaluate Pitt-origin alongside several widely used Pitt-derived datasets under both in-domain and cross-domain settings, using matched and mismatched speech enhancement conditions and multiple state-of-the-art enhancement methods. Our experiments span diverse AD detection architectures, including acoustic-feature-based models, self-supervised speech representation models based on XLS-R~\cite{ConneauXLSR2020}, and Sensitive Layer Selection (SLS)-based models~\cite{zhang2024audio} built on pretrained speech encoders. We further analyze whether recent LALMs, including Kimi-Audio~\cite{ding2025kimi}, Qwen3-Omni~\cite{xu2025qwen3}, Qwen2-Audio~\cite{chu2024qwen2}, Audio Flamingo 3~\cite{ghosh2026audio}, and ultravox-v0\_5-llama-3\_2-1b~\cite{fixie2025ultravox05llama321b}, exhibit similar sensitivity to preprocessing under both \textit{audio-only} and \textit{audio + transcript} settings. The observed effects remain consistent across acoustic-feature-based models, pretrained speech representation models, and LALMs, suggesting that preprocessing sensitivity is not limited to a specific architecture.

\textbf{Contributions.} This paper makes three main contributions: \textbf{(1)} We provide the first systematic benchmark-level analysis of how preprocessing and dataset curation affect downstream AD detection across widely used Pitt-derived datasets. \textbf{(2)} We show that although speech-enhanced datasets can improve in-domain evaluation performance, they frequently reduce robustness in cross-domain evaluation, even under matched enhancement settings. \textbf{(3)} We demonstrate that these effects persist across multiple modeling paradigms, including acoustic-feature-based models, pretrained speech representation models, and recent LALMs, indicating that preprocessing sensitivity is a broader modeling phenomenon rather than an architecture-specific artifact.


\section{Benchmark Datasets and Related Work}
~\label{Sec:Datasets}

\vspace{-7mm}

\subsection{Datasets}
\label{sec:datasets}
The following are six of the most widely used English datasets for speech-based Alzheimer’s disease detection. The sample sizes of these six datasets are shown in Table~\ref{tab:sample_number}, and all of them are available through DementiaBank~\cite{DementiaBank}. More detailed information about the datasets is shown in Appendix~\ref{appendix: dataset}.

\begin{enumerate}
    \item \textbf{Pitt-origin}: 
    Pitt-origin~\cite{BeckerArchNeurol1994} contains four tasks, including the ``Cookie Theft'' picture description task~\cite{goodglass1983boston} for all participants, and three tasks only for participants with AD: semantic fluency, sentence construction, and story recall. The ``Cookie Theft'' picture description task requires participants to orally describe the content of the picture. Since this task is based on spontaneous speech and contains language samples from both participants with AD and healthy controls, it has been widely used.

    \item \textbf{Pitt}: 
    Pitt is a denoised version of Pitt-origin~\cite{BeckerArchNeurol1994} with the same recordings and sample size. However, the speech enhancement pipeline used to construct the dataset was not publicly released. 

    \item \textbf{ADReSS}: 
    ADReSS~\cite{ADReSS} was released as part of the ADReSS Challenge held at Interspeech 2020. All speech samples are derived from Pitt-origin~\cite{BeckerArchNeurol1994}, and have undergone speech enhancement processing (the enhancement method was not released). The dataset was also matched in terms of age and gender.

    \item \textbf{ADReSSo}: 
    ADReSSo~\cite{ADReSSo} was released as part of the ADReSSo Challenge held at Interspeech 2021. All speech samples are derived from Pitt-origin~\cite{BeckerArchNeurol1994}, and have undergone speech enhancement processing (the enhancement method was not released). The dataset was also matched for age and gender. Compared with ADReSS, ADReSSo expanded the sample size.

    \item \textbf{ADReSS-M}: 
    ADReSS-M~\cite{LuzMultilingualICASSP2023} was released as part of the ADReSS-M Challenge held at ICASSP 2023. This dataset uses the same speech samples as ADReSSo, with the same quantity. In contrast to ADReSSo, ADReSS-M did not undergo any speech preprocessing.

    \item \textbf{English Lu Corpus}: 
    Lu~\cite{DementiaBank} is an English speech dataset independent of Pitt-origin, collected under different recording conditions and from different participant populations. Similar to Pitt-origin, the dataset is based on the ``Cookie Theft'' picture description task and contains spontaneous speech recordings without speech preprocessing.

\end{enumerate}

\vspace{-2mm}
\begin{table}[!htbp]
\centering
\small
\begin{tabular}{lccc}
\toprule
\textbf{Dataset} 
& \textbf{Total} 
& \textbf{AD} 
& \textbf{Control} \\
\midrule
Pitt-origin / Pitt  & 552 & 309 & 243 \\
ADReSS              & 156 & 78  & 78  \\
ADReSSo / ADReSS-M  & 237 & 122 & 115 \\
Lu                  & 56 & 27 & 29 \\
\bottomrule
\end{tabular}
\caption{Sample numbers of datasets.}
\vspace{-3mm}
\label{tab:sample_number}
\end{table}

\subsection{Datasets Speech Quality}

\begin{figure*}[t]
\centering
\includegraphics[width=0.95\textwidth]{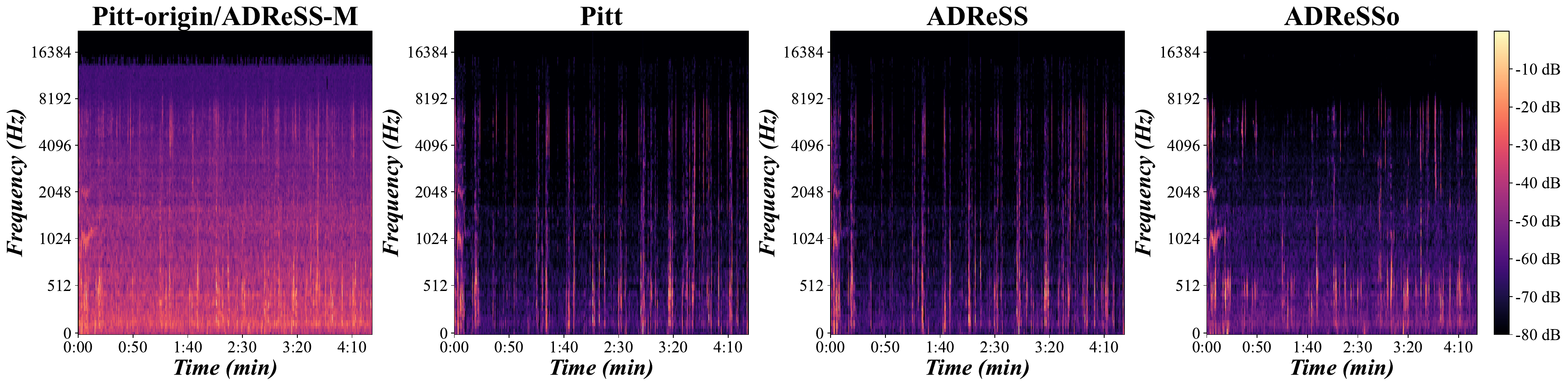}
\caption{Mel spectrograms of the same participant with AD across Pitt-origin and its derived datasets.}
\label{fig:mel}
\end{figure*}

Since Pitt-origin and its four derived datasets originate from the same recordings, we visualize Mel spectrograms for the same participant across all five datasets. As shown in Figure~\ref{fig:mel}, background noise is substantially more prominent in the unprocessed datasets (Pitt-origin and ADReSS-M) than in the denoised datasets (Pitt, ADReSS, and ADReSSo). The denoised datasets exhibit cleaner spectral patterns and reduced background interference, indicating that preprocessing introduces noticeable distributional differences across the datasets. 

To further quantify these differences, we evaluate speech quality of the five datasets using DNSMOS~\cite{reddy2021dnsmos}, a widely used non-intrusive speech quality assessment metric. DNSMOS can evaluate speech samples from three dimensions, including speech quality (SIG), background-noise quality (BAK), and overall quality (OVRL). As shown in Table~\ref{tab:dnsmos_five}, the denoised datasets consistently achieve substantially higher BAK scores than Pitt-origin and ADReSS-M, while SIG scores remain relatively similar across datasets. These results confirm that speech enhancement significantly alters background-noise characteristics.

\begin{table}[!htbp]
\centering
\small
\begin{tabular}{lccc}
\toprule
\textbf{Dataset} 
& \textbf{OVRL}$\uparrow$ 
& \textbf{SIG}$\uparrow$ 
& \textbf{BAK}$\uparrow$ \\
\midrule
Pitt-origin & 2.3073 & 2.9819 & 2.7290 \\
Pitt        & 2.5606 & 3.0479 & 3.6024 \\
ADReSS      & 2.5923 & 3.0857 & 3.5945 \\
ADReSSo     & 2.5222 & 3.0373 & 3.1829 \\
ADReSS-M    & 2.3128 & 2.9805 & 2.6973 \\
\bottomrule
\end{tabular}
\caption{DNSMOS scores of Pitt-origin and its derived datasets. (SIG: speech quality; BAK: background noise quality; OVRL: overall speech quality.)}
\vspace{-2mm}
\label{tab:dnsmos_five}
\end{table}

\subsection{Existing AD Detection Models}
Early AD detection studies primarily explored traditional machine learning approaches. For example, the ADReSS~\cite{ADReSS}, ADReSSo~\cite{ADReSSo} and ADReSS-M~\cite{LuzMultilingualICASSP2023} Challenges adopted support vector machines as a baseline, while Hason et al.~\cite{hason2022spontaneous} used random forests for speech-based AD detection. 

More recently, deep learning has become the dominant paradigm in AD detection. Existing studies have explored hybrid architectures incorporating disfluency-related paralinguistic features~\cite{JinConsenICASSP2023}, end-to-end ASR-based frameworks~\cite{LinICASSP2024}, and transformer-based models for speech-level AD recognition~\cite{DongHAFFormerICASSP2024}. Pu and Zhang et al.~\cite{pu2025integrating} further incorporated pause information into Transformer-based language models for AD detection. Recent work has also begun exploring large language models and large audio-language models for AD detection. For example, Heitz et al.~\cite{heitz2025linguistic} used GPT-4 to extract semantic features from spontaneous speech transcripts, while Park et al.~\cite{park25d_interspeech} applied Chain-of-Thought reasoning and supervised fine-tuning for AD detection.




\section{Methodology}
\subsection{Preprocessing and Distribution Shift}

Speech enhancement and denoising methods inevitably modify the acoustic characteristics of speech signals~\cite{xia2020weighted}. Common effects include enhancement artifacts~\cite{iwamoto2022bad}, spectral smoothing~\cite{li2018conditional}, and other forms of signal distortion. These distortions not only affect the naturalness of speech but may also have a negative impact on downstream models. For example, prior work in automatic speech recognition (ASR) has shown that speech enhancement does not necessarily improve downstream recognition performance and can even reduce robustness under mismatched conditions~\cite{yoshioka2015ntt, chen2018building, menne2019investigation}. 

At the same time, speech enhancement methods, whether based on traditional signal processing or deep learning approaches, can shift the distribution of naturally recorded speech, thereby introducing a significant domain shift~\cite{scheibler24_interspeech, frankel2025automatic}. This issue is particularly relevant for Alzheimer's detection benchmarks. Widely used datasets such as Pitt, ADReSS, and ADReSSo are constructed from Pitt-origin using different preprocessing and curation pipelines, while the exact enhancement procedures were not publicly released. As a result, models trained on these datasets may learn representations associated with processed speech distributions rather than naturally recorded speech. However, real-world clinical speech is typically collected under uncontrolled recording conditions and does not undergo standardized enhancement. 

This mismatch raises an important question: \textit{Do speech enhancement, preprocessing, and dataset curation improve the robustness of AD detection models, or instead introduce distribution shifts that reduce generalization across datasets and recording conditions?}

\subsection{AD Detection Models}
\label{sec:AD model}

\textbf{Traditional Deep Learning Models.} We evaluate enhancement effects across three AD detection architectures with increasing modeling complexity: an acoustic-feature-based eGeMAPS model~\cite{EybenGeMAPS2015}, a self-supervised speech representation model based on XLS-R~\cite{ConneauXLSR2020}, and a Sensitive Layer Selection (SLS)-based model~\cite{zhang2024audio} built on pretrained speech encoders. Model architectures are illustrated in Appendix~\ref{Appendix: Model Architecture}. Among these models, we use the SLS-based model as the primary architecture due to its strong performance in recent speech classification tasks. The model extracts frame-level representations from a frozen pretrained XLS-R encoder and applies Sensitive Layer Selection (SLS)~\cite{zhang2024audio} to learn weighted combinations of Transformer-layer representations, and then performs downstream classification through a classification head.

During the model training phase, we use an 80\%/20\% train-validation split for each dataset with no speaker overlap. Models are trained only on training data under both in-domain and cross-domain settings. For the SLS-based model, we use a learning rate of $3\times10^{-4}$, AdamW optimization, and cross-entropy loss. Training is performed for up to 40 epochs with early stopping patience of 5. All experiments are repeated across five random seeds. All models are trained on NVIDIA RTX 5090 GPUs. Additional implementation details are provided in our released code.

\textbf{Large Audio-Language Models.} To investigate the impact of speech enhancement on large audio-language models (LALMs) in AD detection, we additionally evaluate five recent and widely used LALMs: Kimi-Audio~\cite{ding2025kimi}, Qwen3-Omni~\cite{xu2025qwen3}, Qwen2-Audio~\cite{chu2024qwen2}, Audio Flamingo 3~\cite{ghosh2026audio}, and ultravox-v0\_5-llama-3\_2-1b~\cite{fixie2025ultravox05llama321b}. We first evaluate all models under the zero-shot setting using \textit{audio-only} input. Subsequently, we further evaluate the best-performing Kimi-Audio under zero-shot and few-shot settings using audio-only input and \textit{audio + transcript} input. Prompts are provided in Appendix~\ref{Appendix:Prompt}.

\subsection{Speech Enhancement Methods}
\label{sec:speech enhancement methods}

To evaluate whether all different speech enhancement methods affect AD detection models, we apply five representative speech enhancement models to the downstream AD detection architectures introduced in Section~\ref{sec:AD model}. These models span multiple enhancement paradigms, including waveform-based enhancement (\textit{Denoiser}~\cite{defossez2020real}), complex-domain enhancement (\textit{FRCRN}~\cite{zhao2022frcrn}), Transformer-based enhancement (\textit{MossFormer}~\cite{zhao2023mossformer}), semantic-conditioned enhancement (\textit{Resemble}~\cite{liu2024separate}), and a recent Mamba-based enhancement model (\textit{MAP-SEMamba}~\cite{li2025speech}).

\section{Speech Enhancement Impact on Deep Learning Models}
\label{sec:deep learning test}

We use the three AD detection architectures mentioned in Section~\ref{sec:AD model} to evaluate the impact of preprocessing on AD classification, including an acoustic-feature-based eGeMAPS model, an XLSR-based self-supervised speech model, and an SLS-based model. This section focuses on the SLS-based model, while results for the eGeMAPS and XLS-R models are provided in Appendix~\ref{Appendix: eGeMAPS-based Model} and Appendix~\ref{Appendix: XLSR-based Model}. Across all architectures, the observed preprocessing effects remain broadly consistent, indicating that the phenomenon observed in this paper does not depend on a particular model architecture, but is broadly present.

\subsection{Cross-Dataset Generalization}

To evaluate the impact of speech enhancement on model generalization, we conduct both in-domain and cross-domain experiments on Pitt-origin and its four derived datasets introduced in Section~\ref{Sec:Datasets}. For in-domain evaluation, each dataset is split into 80\% training and 20\% validation sets with no speaker overlap. We use the English Lu Corpus as an independently collected cross-domain test set to evaluate the model’s cross-domain generalization ability.

\begin{figure}[t]
\centering
\includegraphics[width=\columnwidth]{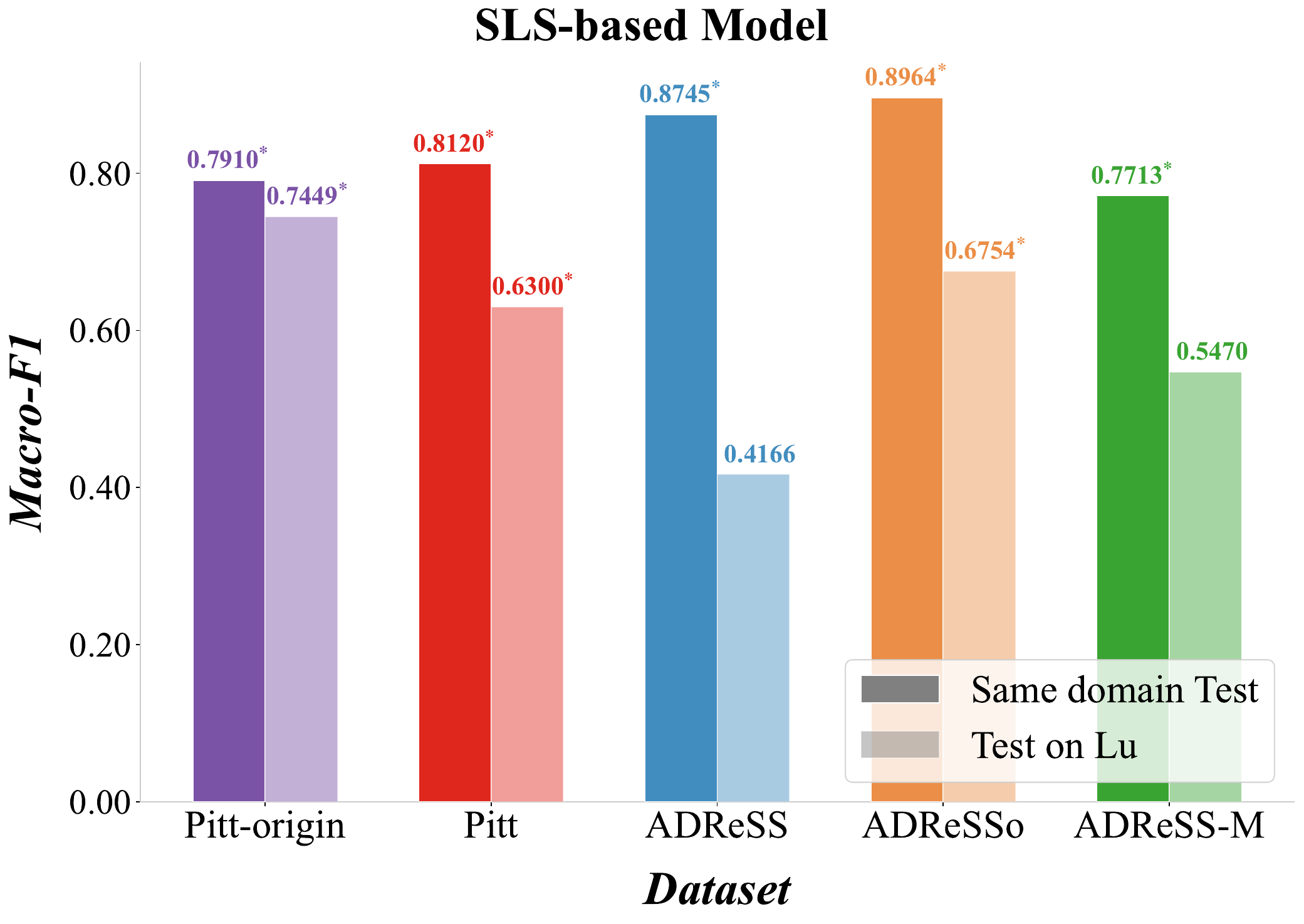}
\caption{SLS-based model generalization test on raw data. 
(\textit{Same Domain Test}: evaluation on the test split of the same dataset used for training; \textit{Test on Lu}: evaluation on the Lu dataset. \textit{Asterisks}$^{*}$: performance significantly different from the binomial majority-class baseline, $p < 0.05$.)}
\vspace{-4mm}
\label{main_content_fig:SLS_unprocessed_generalization}
\end{figure}

As shown in Figure~\ref{main_content_fig:SLS_unprocessed_generalization}, Pitt-origin and its denoised counterpart Pitt achieve comparable in-domain performance, with Pitt performing slightly better. In contrast, ADReSS and ADReSSo further improve in-domain Macro-F1 scores. A possible reason is that speech enhancement, normalization, and dataset curation reduce distributional complexity of the ADReSS and ADReSSo datasets and simplify the learning problem, thereby facilitating in-domain overfitting.

However, this trend changes under cross-domain evaluation. Models trained on the original Pitt-origin data achieve the strongest generalization performance on the Lu corpus, whereas models trained on denoised datasets exhibit clear performance degradation. These results suggest that speech-enhancement-induced distribution shifts reduce cross-dataset generalization. Meanwhile, we found that although ADReSSo achieves relatively strong cross-domain performance on the SLS-based model, this performance is not consistently observed across the eGeMAPS and XLS-R models (Appendix~\ref{Appendix: eGeMAPS Generalization Test On Raw Data}, Appendix~\ref{Appendix: XLSR Generalization Test On Raw Data}), indicating that the effect is architecture-dependent rather than a universal advantage of the dataset itself.

Overall, although speech enhancement and dataset curation may improve in-domain evaluation performance, these gains do not consistently translate to cross-dataset robustness. In contrast, models trained on unprocessed Pitt-origin data exhibit stronger cross-dataset generalization.

\subsection{Speech-Enhanced Test Sets}
\label{sec:4.2}
Using the speech enhancement methods introduced in Section~\ref{sec:speech enhancement methods}, we construct five speech-enhanced versions of the Lu corpus. We train models on Pitt-origin and its four derived datasets (using the 80\% training split of each dataset) and perform cross-domain evaluation on both the original Lu corpus and its enhanced variants.

\begin{figure*}
\centering
\includegraphics[width=0.85\textwidth]{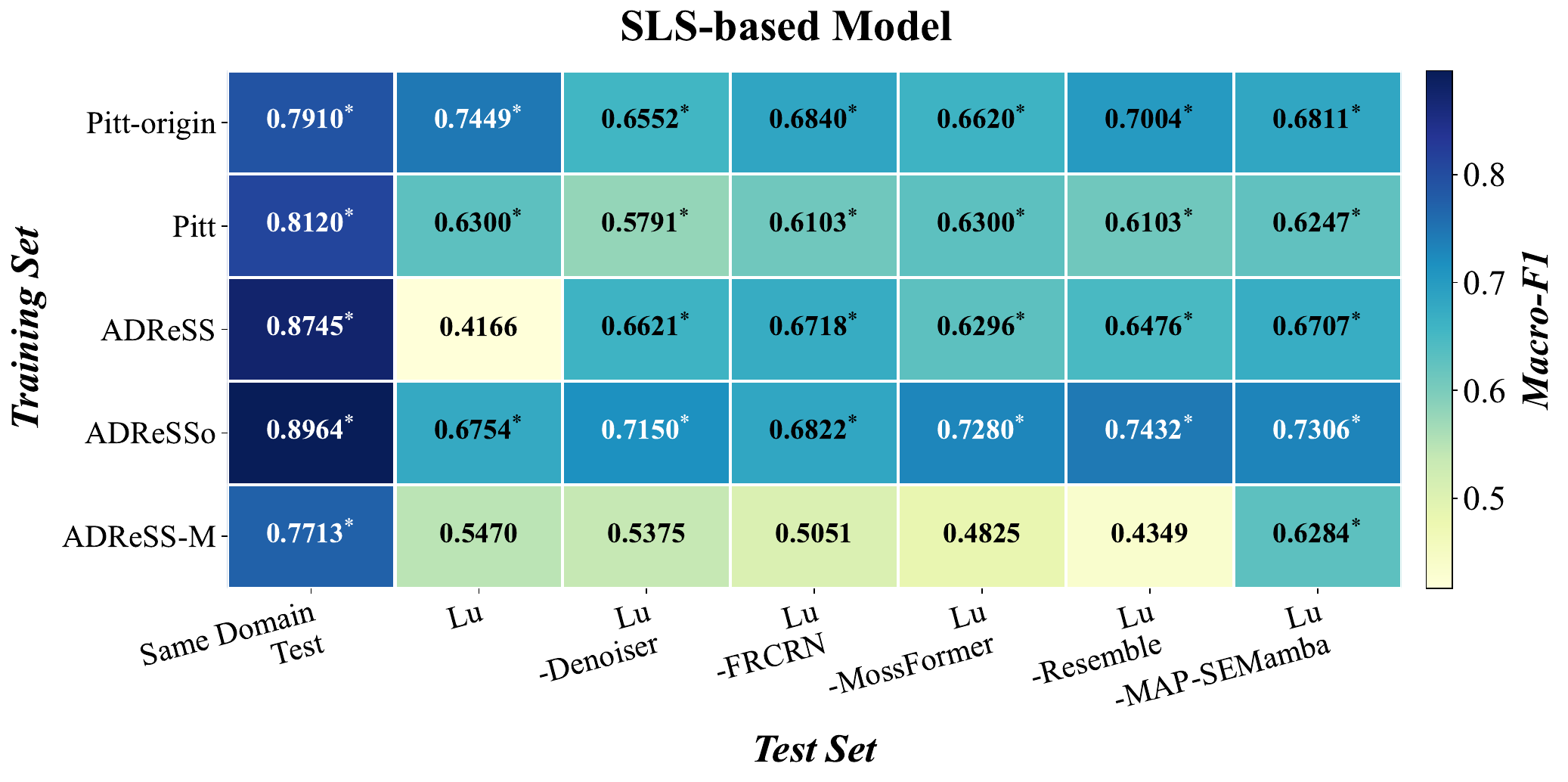}
\vspace{-2mm}
\caption{SLS-based model generalization test on speech-enhanced data. (\textit{Same Domain Test}: evaluation on the test split of the same dataset used for training. \textit{Asterisks}$^{*}$: performance significantly different from the binomial majority-class baseline, $p < 0.05$.)}
\vspace{-3mm}
\label{main_content_fig:generalization}
\end{figure*}

The experimental results are shown in Figure~\ref{main_content_fig:generalization}. The model trained on Pitt-origin achieves the best overall performance on the original unprocessed Lu corpus (Macro-F1 = 0.7449), indicating stronger robustness to unprocessed cross-domain speech. Similar trends are observed for the eGeMAPS and XLS-R models (Appendix~\ref{appendix_fig:xlsr_generalization}, Appendix~\ref{appendix_fig:egemaps_generalization}). However, performance decreases in most cases when the same models are evaluated on speech-enhanced versions of Lu. In contrast, models trained on ADReSS and ADReSSo show clear performance improvements on speech-enhanced Lu test sets. For example, the model trained on ADReSSo achieves a Macro-F1 score of 0.7432 on Lu-Resemble, approaching the best overall result. This suggests that enhanced training data generalize more effectively to similarly enhanced test conditions. Nevertheless, these results still do not surpass the performance of Pitt-origin on the original Lu corpus, and performance varies substantially across different enhancement methods.

Overall, models trained on unprocessed speech perform best on unenhanced test data, whereas models trained on speech-enhanced datasets consistently perform better on similarly enhanced test sets. This suggests that alignment between training and testing enhancement conditions plays an important role in cross-dataset generalization. However, because the preprocessing pipelines used in Pitt, ADReSS, and ADReSSo were not publicly released, it is impossible to reproduce their exact enhancement conditions during evaluation, leading to substantial performance variability across different enhanced test sets.

These findings further raise an important question: \textit{If the same speech enhancement method is applied to both training and test data, can stronger generalization performance be achieved?} To answer this question, we conduct matched enhancement experiments in Section~4.3, where the same speech enhancement methods are applied to both training and testing data.

\subsection{Matched Enhancement Settings}

To determine whether the observed performance degradation is caused primarily by enhancement mismatch between training and test data, we construct a matched enhancement setting. Specifically, we apply five speech enhancement methods introduced in Section~\ref{sec:speech enhancement methods} to the Pitt-origin training data and evaluate models on two Lu test conditions: (1) unprocessed Lu speech (\textit{mismatched}) and (2) Lu speech enhanced using the same enhancement method as the training data (\textit{matched}). We additionally retain the model trained on unprocessed Pitt-origin data as a control (\textit{raw-trained}), with evaluation on the original Lu corpus serving as the \textit{baseline}.

\begin{figure*}
\centering
\includegraphics[width=0.85\textwidth]{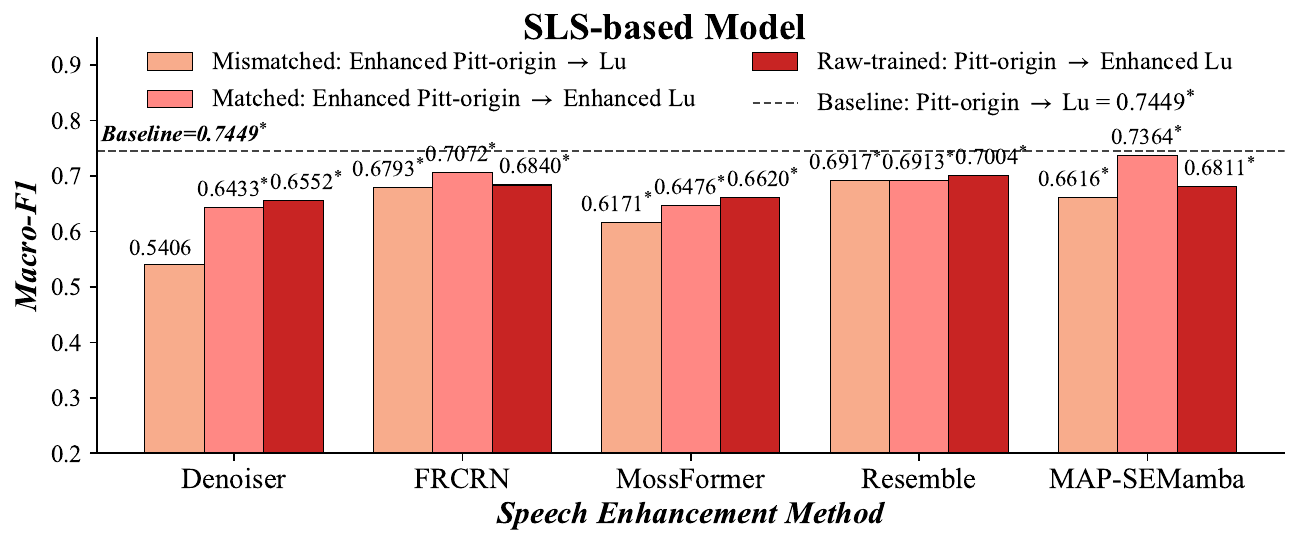}
\vspace{-2mm}
\caption{SLS-based model performance comparison with matched and mismatched speech enhancement methods. (\textit{Mismatched}: enhanced training set with raw test set; 
\textit{Matched}: the same enhancement for training and test set; 
\textit{Raw-trained}: raw training set with enhanced test set; 
\textit{Baseline}: raw training and raw test set. \textit{Asterisks}$^{*}$: performance significantly different from the binomial majority-class baseline, $p < 0.05$.)}
\vspace{-2mm}
\label{main_content_fig:SLS_Match}
\end{figure*}

The experimental results are shown in Figure~\ref{main_content_fig:SLS_Match}. Across all five enhancement methods, the matched setting consistently outperforms the corresponding mismatched setting. These results suggest that alignment between training and testing enhancement conditions partially alleviates enhancement-induced distribution mismatch. However, even under the matched setting, models trained on unprocessed Pitt-origin data (\textit{raw-trained}) still achieve higher Macro-F1 scores in most cases. Moreover, all enhancement-related settings remain below the \textit{baseline}. These results suggest that the observed performance degradation is not caused solely by train-test enhancement mismatch. Instead, speech enhancement itself alters the underlying speech distribution and reduces cross-dataset robustness. Similar trends are observed for the eGeMAPS and XLS-R models (Appendix~\ref{appendix_fig:egemaps_Match} and Appendix~\ref{appendix_fig:XLSR_Match}), indicating that the effect is not limited to a particular model architecture.

Overall, although matched enhancement conditions partially reduce distribution mismatch, they do not fully recover the generalization performance achieved by models trained on unprocessed speech. These findings suggest that speech enhancement itself can substantially affect downstream model robustness in cross-dataset evaluation.

\section{Speech Enhancement Impact on Large Audio-Language Models}

To evaluate the impact of speech enhancement on large audio-language models (LALMs) in Alzheimer’s disease detection, we evaluate five recent and widely used LALMs: Kimi-Audio~\cite{ding2025kimi}, Qwen3-Omni~\cite{xu2025qwen3}, Qwen2-Audio~\cite{chu2024qwen2}, Audio Flamingo 3~\cite{ghosh2026audio}, and ultravox-v0\_5-llama-3\_2-1b~\cite{fixie2025ultravox05llama321b} under multiple inference settings.

\subsection{Audio-Only Settings}

We first evaluate multiple LALMs on all datasets using audio-only input to analyze model behavior when relying solely on speech signals. The experimental results are shown in Appendix~\ref{Appendix: LLMs}.  The results show that most models struggle to reliably distinguish AD from Control under zero-shot audio-only evaluation, while only Kimi-Audio consistently achieves statistically significant performance across most datasets. For example, Qwen2-Audio tends to classify nearly all speech samples as Control, whereas Audio Flamingo 3 tends to classify nearly all samples as AD. In contrast, Kimi-Audio achieves the strongest overall performance and remains statistically significant on all datasets except the Lu corpus.

Based on the above results, we select Kimi-Audio as the main model for further experiments and analysis. We evaluate Kimi-Audio under two inference settings, zero-shot and two-shot. Here, two-shot means that two labeled examples are provided at the inference stage, one from the AD class and one from the control class, without any parameter update or fine-tuning of the model.

\begin{table}[htbp]
\centering
\footnotesize
\begin{tabular}{l c c c}
\toprule
\multirow{2}{*}{\textbf{Test Set}} 
& \multirow{2}{*}{\textbf{Macro-F1}} 
& \textbf{F1 Score} 
& \textbf{F1 Score} \\
& & \textbf{(AD$^{+}$)} & \textbf{(Control$^{+}$)} \\

\midrule
\multicolumn{4}{c}{\textit{Zero-Shot}} \\
\midrule
Pitt-origin & 0.6666$^{*}$ & 0.6726$^{*}$ & 0.6605$^{*}$ \\
Pitt        & 0.5924$^{*}$ & 0.7300$^{*}$ & 0.4548$^{*}$ \\
ADReSS      & 0.5629$^{*}$ & 0.6834$^{*}$ & 0.4425$^{*}$ \\
ADReSSo     & 0.6348$^{*}$ & 0.7273$^{*}$ & 0.5424$^{*}$ \\
ADReSS-M    & 0.7000$^{*}$ & 0.6899$^{*}$ & 0.7102$^{*}$ \\
\midrule
\multicolumn{4}{c}{\textit{Two-Shot}} \\
\midrule
Pitt-origin & 0.6251$^{*}$ & 0.7198$^{*}$ & 0.5305$^{*}$ \\
Pitt        & 0.5878$^{*}$ & 0.7409$^{*}$ & 0.4348$^{*}$ \\
ADReSS      & 0.5879$^{*}$ & 0.6842$^{*}$ & 0.4915$^{*}$ \\
ADReSSo     & 0.6503$^{*}$ & 0.7291$^{*}$ & 0.5714$^{*}$ \\
ADReSS-M    & 0.6441$^{*}$ & 0.5941$^{*}$ & 0.6941$^{*}$ \\
\bottomrule
\end{tabular}
\caption{Kimi-Audio zero-shot and two-shot performance in the audio-only scenario. (\textit{AD$^{+}$ / Control$^{+}$}: F1 score with AD/Control as positive. \textit{Asterisks}$^{*}$: performance significantly different from the binomial majority-class baseline, $p < 0.05$.)}

\label{tab:Audio-only}
\end{table}

The experimental results of Kimi-Audio are shown in Table~\ref{tab:Audio-only}. On the unprocessed Pitt-origin and ADReSS-M datasets, the model produces relatively balanced predictions across AD and Control classes. This suggests that when the input speech is closer to the distribution of unprocessed speech, the model is more likely to rely on relatively stable acoustic patterns. In contrast, on the speech-enhanced datasets (Pitt, ADReSS, and ADReSSo), the model exhibits clear class imbalance, with substantially higher F1 scores with AD as positive than F1 scores with control as positive. These results suggest that speech enhancement substantially alters LALM prediction behavior under audio-only evaluation. This trend remains consistent under both zero-shot and two-shot settings, suggesting that the observed bias is not resolved by providing a small number of labeled examples.

Overall, the audio-only experiments suggest that speech enhancement does not consistently improve the discriminative ability of LALMs and may instead introduce systematic prediction bias, thereby reducing robustness across datasets.

\subsection{Audio + Transcript Settings}

The previous results raise a natural question: \textit{Can textual information mitigate the preprocessing sensitivity observed in audio-only evaluation?} To investigate this, we further evaluate the audio + transcript input format on Pitt-origin and Pitt using Kimi-Audio under both zero-shot and two-shot inference settings.

The results are shown in Table~\ref{tab:Audio+Transcript}. When using audio + transcript input, a trend consistent with the audio-only setting can still be observed. On the denoised Pitt dataset, the model exhibits substantial class imbalance, which becomes even more pronounced under some settings. In contrast, predictions on the unprocessed Pitt-origin dataset remain more balanced across AD and Control classes. These results suggest that adding transcript information does not eliminate enhancement-induced prediction bias. In addition, the two-shot setting also fails to substantially improve this issue. On Pitt, the model continues to exhibit a strong bias toward the AD class. This suggests that simply providing a small number of labeled examples cannot effectively correct the prediction shift caused by changes in data distribution.

\begin{table}[htbp]
\centering
\footnotesize
\begin{tabular}{l c c c}
\toprule
\multirow{2}{*}{\textbf{Test Set}} 
& \multirow{2}{*}{\textbf{Macro-F1}} 
& \textbf{F1 Score} 
& \textbf{F1 Score} \\
& & \textbf{(AD$^{+}$)} & \textbf{(Control$^{+}$)} \\
\midrule
\multicolumn{4}{c}{\textit{Zero-Shot}} \\
\midrule
Pitt-origin & 0.6514$^{*}$ & 0.6350$^{*}$ & 0.6678$^{*}$ \\
Pitt        & 0.4756 & 0.7164 & 0.2348 \\
\midrule
\multicolumn{4}{c}{\textit{Two-Shot}} \\
\midrule
Pitt-origin & 0.6651$^{*}$ & 0.7313$^{*}$ & 0.5988$^{*}$ \\
Pitt        & 0.5498$^{*}$ & 0.7194$^{*}$ & 0.3801$^{*}$ \\
\bottomrule
\end{tabular}
\caption{Kimi-Audio zero-shot and two-shot performance in audio + transcript scenario. (\textit{AD$^{+}$ / Control$^{+}$}: F1 score with AD/Control as positive. \textit{Asterisks}$^{*}$: performance significantly different from the binomial majority-class baseline, $p < 0.05$.)}
\vspace{-3mm}
\label{tab:Audio+Transcript}
\end{table}

Overall, the multimodal experiments further suggest that speech enhancement substantially affects downstream LALM prediction behavior, even when textual information is available. 

\section{Conclusion}
This paper revisits the role of speech preprocessing and dataset curation in speech-based Alzheimer’s disease (AD) detection across widely used Pitt-derived benchmarks. We systematically evaluate the effects of speech enhancement using multiple experimental settings, including in-domain and cross-domain evaluation, speech-enhanced test conditions, matched and mismatched enhancement settings, and multiple speech enhancement pipelines. We further analyze whether these effects remain consistent across different modeling paradigms, including acoustic-feature-based models, pretrained speech representation models, and recent large audio-language models. Across both supervised AD detection models and LALMs, the results consistently show that speech-enhanced datasets often improve in-domain evaluation performance while reducing cross-dataset robustness. Even when matched enhancement conditions are applied to both training and test data, models trained on unprocessed Pitt-origin speech generally achieve stronger generalization performance. In addition, LALMs exhibit systematic prediction bias under speech-enhanced conditions, suggesting that preprocessing sensitivity extends beyond conventional supervised architectures to modern foundation audio models.

Overall, our findings show that speech enhancement and dataset curation are not merely preprocessing choices, but factors that can substantially influence downstream AD detection behavior and cross-dataset generalization. These results highlight the importance of carefully considering preprocessing pipelines when constructing and evaluating speech-based AD detection benchmarks, and suggest that cleaner speech does not necessarily lead to more robust AD detection systems.

\section{Limitations}
Our research focuses on Alzheimer’s disease speech detection from speech in English contexts. In this field, the vast majority of studies have used the Pitt Corpus~\cite{BeckerArchNeurol1994} and its derived datasets, and the number of publicly available English AD speech datasets independent of the Pitt Corpus is very limited. We note that although there are some larger English AD speech datasets, these datasets have not yet been made publicly available and therefore cannot be used for experimental validation in this study. Other publicly accessible English dementia datasets, such as the VAS Corpus~\cite{liang2022evaluating} and the English WLS Corpus~\cite{herd2014cohort}, are not restricted to AD patients, as they include individuals with MCI or other forms of cognitive decline without distinguishing these diagnostic groups. To ensure the rigor of the research, we did not use these datasets in our experiments. Therefore, due to the limited number of publicly available English AD speech datasets independent of the Pitt Corpus, this paper selects the English Lu Corpus~\cite{DementiaBank} as the cross-domain test set to evaluate the model’s generalization ability.

Our research primarily focuses on the impact of speech enhancement on speech-based AD detection models, with particular attention to the relationship between speech enhancement processing and model generalization ability. Therefore, we only use text as an auxiliary diagnostic tool for LALMs and do not conduct an in-depth study on using text to detect AD.

Although our experimental results consistently show that, compared with Pitt, ADReSS, ADReSSo, and ADReSS-M, models trained on the original Pitt-origin generally have stronger generalization ability, it should be particularly noted that Pitt-origin itself also contains substantial environmental noise. How to improve the robustness of Alzheimer’s disease detection models to noise remains an important and unresolved problem. Future work needs to explore more robust model architectures, enabling models to stably extract Alzheimer’s disease-related pathological features from noisy speech.


\section*{Use of AI Assistants}
We used GPT-5.5 and Claude Opus 4.7 for code assistance, and the same models purely for language-related assistance in writing the paper.

\section*{Licensing}
We use the DementiaBank dataset, a shared multimedia database for studying communication in dementia. DementiaBank is part of TalkBank. Therefore, use of the data is generally governed by the Creative Commons Attribution–NonCommercial–ShareAlike 3.0 license (CC BY-NC-SA 3.0), unless otherwise specified. This license requires attribution, restricts commercial use, and requires share-alike for derivatives. Access to DementiaBank is password-protected and restricted to approved members of the DementiaBank consortium. We comply with the DementiaBank data use requirements, including restrictions on redistributing or posting password-protected data on external websites or servers.

\bibliography{custom}

@article{AlzheimerEurope2019,
  author={Europe, Alzheimer},
  title={{Dementia in Europe Yearbook 2019: Estimating the prevalence of dementia in Europe}},
  journal={Alzheimer Europe},
  volume={180},
  year={2019}
}

@article{AlzheimersFactsFigures2023,
  title={{Alzheimer’s disease facts and figures}},
  author={Better, MAPPING A},
  journal={Alzheimers Dement},
  volume={19},
  number={4},
  pages={1598--1695},
  year={2023}
}

@inproceedings{ADReSS,
  title     = {{Alzheimer’s Dementia Recognition Through Spontaneous Speech: The ADReSS Challenge}},
  author    = {Saturnino Luz and Fasih Haider and Sofia de la Fuente and Davida Fromm and Brian MacWhinney},
  year      = {2020},
  booktitle = {Interspeech},
  pages     = {2172--2176},
}

@inproceedings{ADReSSo,
title = {Detecting Cognitive Decline Using Speech Only: The ADReSSo Challenge},
author = {Saturnino Luz and Fasih Haider and {Sofia de la Fuente} and Davida Fromm and Brian MacWhinney},
year = {2021},
booktitle = {Interspeech},
pages = {3780--3784}
}

@inproceedings{LuzMultilingualICASSP2023,
  title={{Multilingual alzheimer’s dementia recognition through spontaneous speech: a signal processing grand challenge}},
  author={Luz, Saturnino and Haider, Fasih and Fromm, Davida and Lazarou, Ioulietta and Kompatsiaris, Ioannis and MacWhinney, Brian},
  booktitle={International Conference on Acoustics, Speech and Signal Processing (ICASSP)},
  pages={1--2},
  year={2023},
  organization={IEEE}
}

@article{BeckerArchNeurol1994,
  title={{The natural history of Alzheimer's disease: description of study cohort and accuracy of diagnosis}},
  author={Becker, James T and Boiler, Fran{\c{c}}ois and Lopez, Oscar L and Saxton, Judith and McGonigle, Karen L},
  journal={Archives of neurology},
  volume={51},
  number={6},
  pages={585--594},
  year={1994},
  publisher={American Medical Association}
}

@book{goodglass1983boston,
  title={Boston diagnostic aphasia examination booklet},
  author={Goodglass, Harold and Kaplan, Edith},
  year={1983},
  publisher={Lea \& Febiger}
}

@article{DementiaBank,
  title={{DementiaBank: Theoretical rationale, protocol, and illustrative analyses}},
  author={Lanzi, Alyssa M and Saylor, Anna K and Fromm, Davida and Liu, Houjun and MacWhinney, Brian and Cohen, Matthew L},
  journal={American Journal of Speech-Language Pathology},
  volume={32},
  number={2},
  pages={426--438},
  year={2023},
  publisher={American Speech-Language-Hearing Association}
}

@inproceedings{frankel2025automatic,
  title={Automatic Detection of Domain Shifts in Speech Enhancement Systems Using Confidence-Based Metrics},
  author={Frankel, Lior and Chazan, Shlomo E and Goldberger, Jacob},
  booktitle={ICASSP 2025-2025 IEEE International Conference on Acoustics, Speech and Signal Processing (ICASSP)},
  pages={1--5},
  year={2025},
  organization={IEEE}
}

@inproceedings{xia2020weighted,
  title={Weighted speech distortion losses for neural-network-based real-time speech enhancement},
  author={Xia, Yangyang and Braun, Sebastian and Reddy, Chandan KA and Dubey, Harishchandra and Cutler, Ross and Tashev, Ivan},
  booktitle={ICASSP 2020-2020 IEEE International Conference on Acoustics, Speech and Signal Processing (ICASSP)},
  pages={871--875},
  year={2020},
  organization={IEEE}
}

@inproceedings{iwamoto2022bad,
  title     = {{How bad are artifacts?: Analyzing the impact of speech enhancement errors on ASR}},
  author    = {Kazuma Iwamoto and Tsubasa Ochiai and Marc Delcroix and Rintaro Ikeshita and Hiroshi Sato and Shoko Araki and Shigeru Katagiri},
  year      = {2022},
  booktitle = {{Interspeech 2022}},
  pages     = {5418--5422}
}

@article{li2018conditional,
  title={A conditional generative model for speech enhancement},
  author={Li, Zeng-Xi and Dai, Li-Rong and Song, Yan and McLoughlin, Ian},
  journal={Circuits, Systems, and Signal Processing},
  volume={37},
  number={11},
  pages={5005--5022},
  year={2018},
  publisher={Springer}
}

@inproceedings{yoshioka2015ntt,
  title={The NTT CHiME-3 system: Advances in speech enhancement and recognition for mobile multi-microphone devices},
  author={Yoshioka, Takuya and Ito, Nobutaka and Delcroix, Marc and Ogawa, Atsunori and Kinoshita, Keisuke and Fujimoto, Masakiyo and Yu, Chengzhu and Fabian, Wojciech J and Espi, Miquel and Higuchi, Takuya and others},
  booktitle={2015 IEEE Workshop on Automatic Speech Recognition and Understanding (ASRU)},
  pages={436--443},
  year={2015},
  organization={IEEE}
}

@inproceedings{chen2018building,
  title     = {{Building State-of-the-art Distant Speech Recognition Using the CHiME-4 Challenge with a Setup of Speech Enhancement Baseline}},
  author    = {Szu-Jui Chen and Aswin Shanmugam Subramanian and Hainan Xu and Shinji Watanabe},
  year      = {2018},
  booktitle = {{Interspeech 2018}},
  pages     = {1571--1575}
}

@inproceedings{menne2019investigation,
  title={Investigation into joint optimization of single channel speech enhancement and acoustic modeling for robust ASR},
  author={Menne, Tobias and Schl{\"u}ter, Ralf and Ney, Hermann},
  booktitle={ICASSP 2019-2019 IEEE International Conference on Acoustics, Speech and Signal Processing (ICASSP)},
  pages={6660--6664},
  year={2019},
  organization={IEEE}
}

@article{Scheltens2021,
  title={{Alzheimer's disease}},
  author={Scheltens, Philip and De Strooper, Bart and Kivipelto, Miia and Holstege, Henne and Ch{\'e}telat, Gael and Teunissen, Charlotte E and Cummings, Jeffrey and van der Flier, Wiesje M},
  journal={The Lancet},
  volume={397},
  number={10284},
  pages={1577--1590},
  year={2021},
  publisher={Elsevier}
}

@inproceedings{reddy2021dnsmos,
  title={DNSMOS: A non-intrusive perceptual objective speech quality metric to evaluate noise suppressors},
  author={Reddy, Chandan KA and Gopal, Vishak and Cutler, Ross},
  booktitle={ICASSP 2021-2021 IEEE International Conference on Acoustics, Speech and Signal Processing (ICASSP)},
  pages={6493--6497},
  year={2021},
  organization={IEEE}
}

@inproceedings{defossez2020real,
  title     = {{Real Time Speech Enhancement in the Waveform Domain}},
  author    = {Alexandre Défossez and Gabriel Synnaeve and Yossi Adi},
  year      = {2020},
  booktitle = {{Interspeech 2020}},
  pages     = {3291--3295}
}

@inproceedings{zhao2022frcrn,
  title={FRCRN: Boosting feature representation using frequency recurrence for monaural speech enhancement},
  author={Zhao, Shengkui and Ma, Bin and Watcharasupat, Karn N and Gan, Woon-Seng},
  booktitle={ICASSP 2022-2022 IEEE international conference on acoustics, speech and signal processing (ICASSP)},
  pages={9281--9285},
  year={2022},
  organization={IEEE}
}

@inproceedings{zhao2023mossformer,
  title={Mossformer: Pushing the performance limit of monaural speech separation using gated single-head transformer with convolution-augmented joint self-attentions},
  author={Zhao, Shengkui and Ma, Bin},
  booktitle={ICASSP 2023-2023 IEEE International Conference on Acoustics, Speech and Signal Processing (ICASSP)},
  pages={1--5},
  year={2023},
  organization={IEEE}
}

@article{liu2024separate,
  title={Separate anything you describe},
  author={Liu, Xubo and Kong, Qiuqiang and Zhao, Yan and Liu, Haohe and Yuan, Yi and Liu, Yuzhuo and Xia, Rui and Wang, Yuxuan and Plumbley, Mark D and Wang, Wenwu},
  journal={IEEE Transactions on Audio, Speech and Language Processing},
  volume={33},
  pages={458--471},
  year={2024},
  publisher={IEEE}
}

@inproceedings{ConneauXLSR2020,
  title     = {{Unsupervised Cross-Lingual Representation Learning for Speech Recognition}},
  author    = {Alexis Conneau and Alexei Baevski and Ronan Collobert and Abdelrahman Mohamed and Michael Auli},
  year      = {2021},
  booktitle = {Interspeech},
  pages     = {2426--2430}
}

@article{ding2025kimi,
  title={Kimi-audio technical report},
  author={Ding, Ding and Ju, Zeqian and Leng, Yichong and Liu, Songxiang and Liu, Tong and Shang, Zeyu and Shen, Kai and Song, Wei and Tan, Xu and Tang, Heyi and others},
  journal={arXiv preprint arXiv:2504.18425},
  year={2025}
}

@inproceedings{LinICASSP2024,
  title={{Dementia Assessment Using Mandarin Speech with an Attention-Based Speech Recognition Encoder}},
  author={Lin, Zih-Jyun and Chen, Yi-Ju and Kuo, Po-Chih and Huang, Likai and Hu, Chaur-Jong and Chen, Cheng-Yu},
  booktitle={International Conference on Acoustics, Speech and Signal Processing (ICASSP)},
  pages={12461--12465},
  year={2024},
  organization={IEEE}
}

@inproceedings{DongHAFFormerICASSP2024,
  title={{HAFFormer: A hierarchical attention-free framework for Alzheimer’s disease detection from spontaneous speech}},
  author={Dong, Zhongren and Zhang, Zixing and Xu, Weixiang and Han, Jing and Ou, Jianjun and Schuller, Bj{\"o}rn W},
  booktitle={International Conference on Acoustics, Speech and Signal Processing (ICASSP)},
  pages={11246--11250},
  year={2024},
  organization={IEEE}
}

@inproceedings{JinConsenICASSP2023,
  title={{Consen: Complementary and simultaneous ensemble for alzheimer’s disease detection and mmse score prediction}},
  author={Jin, Longbin and Oh, Yealim and Kim, Hyunseo and Jung, Hyuntaek and Jon, Hyo Jin and Shin, Jung Eun and Kim, Eun Yi},
  booktitle={International Conference on Acoustics, Speech and Signal Processing (ICASSP)},
  pages={1--2},
  year={2023},
  organization={IEEE}
}

@article{literaturereview,
  title={{Deep learning-based speech analysis for Alzheimer’s disease detection: a literature review}},
  author={Yang, Qin and Li, Xin and Ding, Xinyun and Xu, Feiyang and Ling, Zhenhua},
  journal={Alzheimer's Research \& Therapy},
  volume={14},
  number={1},
  pages={186},
  year={2022},
  publisher={Springer}
}

@article{literaturereview1,
  title={{Speech based detection of Alzheimer’s disease: a survey of AI techniques, datasets and challenges}},
  author={Ding, Kewen and Chetty, Madhu and Noori Hoshyar, Azadeh and Bhattacharya, Tanusri and Klein, Britt},
  journal={Artificial Intelligence Review},
  volume={57},
  number={12},
  pages={325},
  year={2024},
  publisher={Springer}
}

@article{EybenGeMAPS2015,
  title={{The Geneva minimalistic acoustic parameter set (GeMAPS) for voice research and affective computing}},
  author={Eyben, Florian and Scherer, Klaus R and Schuller, Bj{\"o}rn W and Sundberg, Johan and Andr{\'e}, Elisabeth and Busso, Carlos and Devillers, Laurence Y and Epps, Julien and Laukka, Petri and Narayanan, Shrikanth S and others},
  journal={IEEE transactions on affective computing},
  volume={7},
  number={2},
  pages={190--202},
  year={2015},
  publisher={IEEE}
}

@inproceedings{EybenOpenSMILE2010,
  title={{Opensmile: the munich versatile and fast open-source audio feature extractor}},
  author={Eyben, Florian and W{\"o}llmer, Martin and Schuller, Bj{\"o}rn},
  booktitle={Proceedings of the 18th ACM international conference on Multimedia},
  pages={1459--1462},
  year={2010},
  publisher={ACM}
}

@inproceedings{zhang2024audio,
  title={{Audio deepfake detection with self-supervised xls-r and sls classifier}},
  author={Zhang, Qishan and Wen, Shuangbing and Hu, Tao},
  booktitle={Proceedings of the 32nd ACM International Conference on Multimedia},
  pages={6765--6773},
  year={2024}
}

@article{xu2025qwen3,
  title={Qwen3-omni technical report},
  author={Xu, Jin and Guo, Zhifang and Hu, Hangrui and Chu, Yunfei and Wang, Xiong and He, Jinzheng and Wang, Yuxuan and Shi, Xian and He, Ting and Zhu, Xinfa and others},
  journal={arXiv preprint arXiv:2509.17765},
  year={2025}
}

@article{chu2024qwen2,
  title={Qwen2-audio technical report},
  author={Chu, Yunfei and Xu, Jin and Yang, Qian and Wei, Haojie and Wei, Xipin and Guo, Zhifang and Leng, Yichong and Lv, Yuanjun and He, Jinzheng and Lin, Junyang and others},
  journal={arXiv preprint arXiv:2407.10759},
  year={2024}
}

@article{ghosh2026audio,
  title={Audio flamingo 3: Advancing audio intelligence with fully open large audio language models},
  author={Ghosh, Sreyan and Goel, Arushi and Kim, Jaehyeon and Kumar, Sonal and Kong, Zhifeng and Lee, Sang-gil and Yang, Chao-Han and Duraiswami, Ramani and Manocha, Dinesh and Valle, Rafael and others},
  journal={Advances in Neural Information Processing Systems},
  volume={38},
  pages={41819--41886},
  year={2026}
}

@misc{fixie2025ultravox05llama321b,
  title        = {Ultravox v0.5 Llama 3.2 1B},
  author       = {{Fixie.ai}},
  year         = {2025},
  howpublished = {\url{https://huggingface.co/fixie-ai/ultravox-v0_5-llama-3_2-1b}},
  note         = {Hugging Face model card}
}

@article{VaswaniTransformer2017,
  title={{Attention is all you need}},
  author={Vaswani, Ashish and Shazeer, Noam and Parmar, Niki and Uszkoreit, Jakob and Jones, Llion and Gomez, Aidan N and Kaiser, {\L}ukasz and Polosukhin, Illia},
  journal={Advances in neural information processing systems},
  volume={30},
  year={2017},
  publisher={ACM}
}

@inproceedings{li2025speech,
  title={Speech Enhancement with MAP-based Training for Robust ASR},
  author={Li, You-Jin and Chao, Rong and Su, Borching and Tsao, Yu},
  booktitle={ICASSP 2025-2025 IEEE International Conference on Acoustics, Speech and Signal Processing (ICASSP)},
  pages={1--5},
  year={2025},
  organization={IEEE}
}

@inproceedings{heitz2025linguistic,
  title={Linguistic features extracted by GPT-4 improve Alzheimer’s disease detection based on spontaneous speech},
  author={Heitz, Jonathan and Schneider, Gerold and Langer, Nicolas},
  booktitle={Proceedings of the 31st International Conference on Computational Linguistics},
  pages={1850--1864},
  year={2025}
}

@inproceedings{pu2025integrating,
  title={Integrating pause information with word embeddings in language models for Alzheimer’s disease detection from spontaneous speech},
  author={Pu, Yu and Zhang, Wei-Qiang},
  booktitle={ICASSP 2025-2025 IEEE International Conference on Acoustics, Speech and Signal Processing (ICASSP)},
  pages={1--5},
  year={2025},
  organization={IEEE}
}

@article{hason2022spontaneous,
  title={Spontaneous speech feature analysis for alzheimer's disease screening using a random forest classifier},
  author={Hason, Lior and Krishnan, Sri},
  journal={Frontiers in Digital Health},
  volume={4},
  pages={901419},
  year={2022},
  publisher={Frontiers Media SA}
}

@inproceedings{park25d_interspeech,
  title     = {{Reasoning-Based Approach with Chain-of-Thought for Alzheimer’s Detection Using Speech and Large Language Models}},
  author    = {Chanwoo Park and Anna Seo Gyeong Choi and Sunghye Cho and Chanwoo Kim},
  year      = {2025},
  booktitle = {{Interspeech 2025}},
  pages     = {2185--2189}
}

@article{liang2022evaluating,
  title={Evaluating voice-assistant commands for dementia detection},
  author={Liang, Xiaohui and Batsis, John A and Zhu, Youxiang and Driesse, Tiffany M and Roth, Robert M and Kotz, David and MacWhinney, Brian},
  journal={Computer Speech \& Language},
  volume={72},
  pages={101297},
  year={2022},
  publisher={Elsevier}
}

@article{herd2014cohort,
  title={Cohort profile: Wisconsin longitudinal study (WLS)},
  author={Herd, Pamela and Carr, Deborah and Roan, Carol},
  journal={International journal of epidemiology},
  volume={43},
  number={1},
  pages={34--41},
  year={2014},
  publisher={Oxford University Press}
}

@inproceedings{scheibler24_interspeech,
  title     = {{Universal Score-based Speech Enhancement with High Content Preservation}},
  author    = {Robin Scheibler and Yusuke Fujita and Yuma Shirahata and Tatsuya Komatsu},
  year      = {2024},
  booktitle = {{Interspeech 2024}},
  pages     = {1165--1169}
}

\appendix

\clearpage
\section{Dataset Information}
\label{appendix: dataset}

\subsection{Dataset Details}

\begin{table}[!htbp]
\centering
\begin{tabular}{c|cc}
\hline
 & \textbf{Non-Denoising} & \textbf{Denoised} \\
\hline
\textbf{Original Size} & Pitt-origin & Pitt \\
\textbf{Reduced Size} & ADReSS-M & \begin{tabular}[c]{@{}c@{}}ADReSSo, \\ ADReSS\end{tabular} \\
\hline
\end{tabular}
\caption{Relationship of Pitt-origin and its derived datasets. (Pitt-origin and Pitt have the same size. ADReSS-M and ADReSSo have the same size. ADReSS is smaller than ADReSS-M and ADReSSo.)}
\label{appendix_tab:dataset relationship}
\end{table}

\begin{figure}[!htbp]
\centering
\includegraphics[width=0.9\columnwidth]{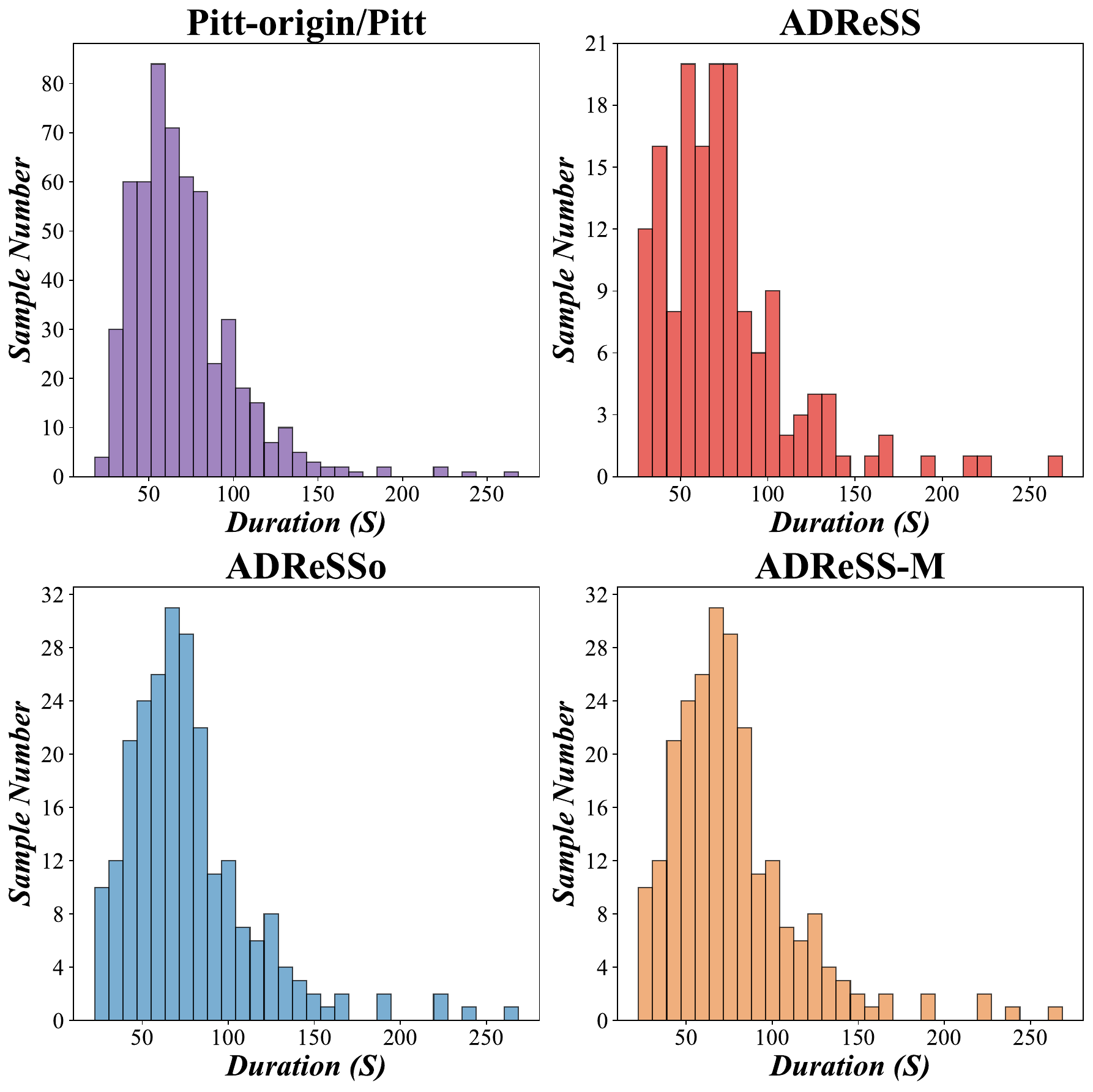}
\caption{Duration distribution of Pitt-origin and its derived datasets.}
\vspace{-2mm}
\label{appendix_fig:duration distribution}
\end{figure}

\begin{table}[!htbp]
\centering
\small
\hspace*{-3mm}
\begin{tabular}{lccccc}
\toprule
\textbf{Dataset} 
& \textbf{Min(s)} & \textbf{Max(s)} & \textbf{Mean(s)} & \textbf{Total(min)} \\
\midrule
Pitt-origin / Pitt  & 17.89 & 268.49 & 69.97 & 643.72 \\
ADReSS              & 26.06 & 268.49 & 75.30 & 195.78 \\
ADReSSo             & 22.35 & 268.49 & 76.89 & 303.72 \\
ADReSS-M            & 22.35 & 268.49 & 76.89 & 303.72 \\
\bottomrule
\end{tabular}
\caption{Duration distribution of Pitt-origin and its derived datasets.}
\label{tab:dataset_duration}
\end{table}

\setlength{\cmidrulewidth}{0.4pt}

\begin{table}[!htbp]
\centering
\small
\begin{tabular}{lcccc}
\toprule
\multirow{2}{*}{Age} & \multicolumn{2}{c}{AD} & \multicolumn{2}{c}{Control} \\
\cline{2-5}
 & Male & Female & Male & Female \\
\hline
$<50$ & 0 & 1 & 0 & 6 \\
$[50, 55)$ & 5 & 1 & 8 & 9 \\
$[55, 60)$ & 10 & 11 & 17 & 28 \\
$[60, 65)$ & 18 & 18 & 13 & 25 \\
$[65, 70)$ & 26 & 33 & 24 & 40 \\
$[70, 75)$ & 23 & 43 & 20 & 33 \\
$[75, 80)$ & 23 & 45 & 6 & 9 \\
$[80, 90]$ & 13 & 37 & 1 & 4 \\
unknown & 2 & 0 & 0 & 0 \\
\hline
Total & 120 & 189 & 89 & 154 \\
\bottomrule
\end{tabular}
\caption{Pitt-origin / Pitt dataset age and gender metadata.}
\end{table}

\begin{table}[!htbp]
\centering
\small
\begin{tabular}{lcccc}
\toprule
\multirow{2}{*}{Age} & \multicolumn{2}{c}{AD} & \multicolumn{2}{c}{Control} \\
\cline{2-5}
 & Male & Female & Male & Female \\
\hline
$[50, 55)$ & 2 & 0 & 2 & 0 \\
$[55, 60)$ & 7 & 6 & 7 & 6 \\
$[60, 65)$ & 4 & 9 & 4 & 9 \\
$[65, 70)$ & 9 & 14 & 9 & 14 \\
$[70, 75)$ & 9 & 11 & 9 & 11 \\
$[75, 80)$ & 4 & 3 & 4 & 3 \\
\hline
Total & 35 & 43 & 35 & 43 \\
\bottomrule
\end{tabular}
\caption{ADReSS dataset age and gender metadata.}
\end{table}

\begin{table}[!htbp]
\centering
\small
\begin{tabular}{lcccc}
\toprule
\multirow{2}{*}{Age} & \multicolumn{2}{c}{AD} & \multicolumn{2}{c}{Control} \\
\cline{2-5}
 & Male & Female & Male & Female \\
\hline
$[50, 55)$ & 1 & 0 & 1 & 1 \\
$[55, 60)$ & 7 & 6 & 6 & 16 \\
$[60, 65)$ & 7 & 8 & 7 & 13 \\
$[65, 70)$ & 7 & 22 & 11 & 23 \\
$[70, 75)$ & 9 & 21 & 10 & 18 \\
$[75, 80]$ & 12 & 22 & 5 & 4 \\
\hline
Total & 43 & 79 & 40 & 75 \\
\bottomrule
\end{tabular}
\caption{ADReSSo / ADReSS-M dataset age and gender metadata.}
\end{table}

\setlength{\cmidrulewidth}{0.4pt}

\begin{table}[!htbp]
\centering
\tiny
\resizebox{\columnwidth}{!}{
\begin{tabular}{lccc}
\toprule
\textbf{Dataset} & \textbf{OVRL}$\uparrow$ & \textbf{SIG}$\uparrow$ & \textbf{BAK}$\uparrow$ \\
\midrule
Pitt-origin            & 2.3073 & 2.9819 & 2.7290 \\
Pitt-origin-Denoiser   & 2.9687 & 3.2946 & 4.2052 \\
Pitt-origin-FRCRN      & 2.8704 & 3.3260 & 3.9332 \\
Pitt-origin-MossFormer & 3.0232 & 3.4136 & 4.1258 \\
Pitt-origin-Resemble   & 2.8856 & 3.3151 & 3.8913 \\
Pitt-origin-MAP-SEMamba  & 2.8799 & 3.3434 & 3.8376 \\
\bottomrule
\end{tabular}
}
\caption{DNSMOS scores of Pitt-origin under different speech enhancement methods. (SIG: speech quality; BAK: background noise quality; OVRL: overall speech quality.)}
\label{tab:DNSMOS}
\end{table}

\clearpage
\onecolumn
\section{Prompt}
\label{Appendix:Prompt}

\subsection{Audio Prompt}
\label{Appendix:audio_Prompt}

\begin{tcolorbox}[colback=gray!10,colframe=black]
\ttfamily\small
You are a clinical speech-language pathologist specialized in detecting Alzheimer's disease and dementia from spontaneous speech. You analyze speech patterns including: word-finding difficulties, semantic paraphasias, empty speech, reduced syntactic complexity, repetitions, incomplete utterances, and pragmatic impairments. Based on the audio, classify the speaker. Listen to this speech sample carefully. Based on the speech characteristics, is this speaker showing signs of dementia or is this a healthy control? Answer with exactly one word: 'Dementia' or 'Control'.
\end{tcolorbox}

\subsection{Audio + Transcript Prompt}

\begin{tcolorbox}[colback=gray!10,colframe=black]
\ttfamily\small
You are a clinical speech-language pathologist specialized in detecting Alzheimer's disease and dementia from spontaneous speech and transcription. You analyze speech patterns and transcribed text including: word-finding difficulties, semantic paraphasias, empty speech, reduced syntactic complexity, repetitions, incomplete utterances, and pragmatic impairments. Based on the audio and transcription, classify the speaker. Listen to this speech sample and read the transcription carefully. Based on both the speech characteristics and the transcription content, is this speaker showing signs of dementia or is this a healthy control? Answer with exactly one word: 'Dementia' or 'Control'.
\end{tcolorbox}

\clearpage
\onecolumn
\section{Large Audio-Language Models}
~\label{Appendix: LLMs}
\vspace{-8mm}

\subsection{LALMs Audio-only Zero-Shot Setting}
\vspace{-2mm}

\begin{figure}[H]
\centering
\includegraphics[width=0.9\textwidth]{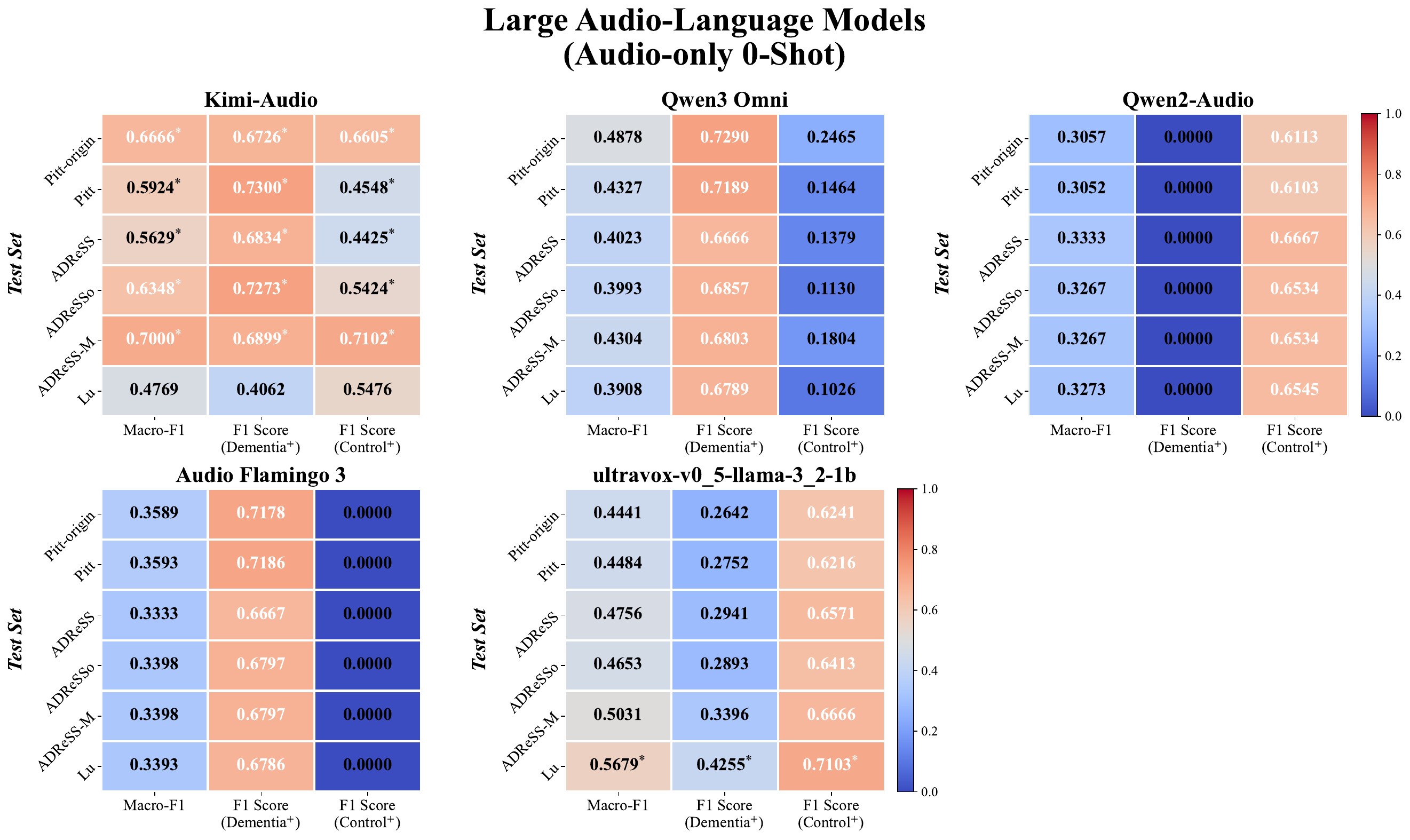}
\vspace{-3mm}
\caption{LALMs performance in the audio-only zero-shot setting. (\textit{AD$^{+}$ / Control$^{+}$}: F1 score with AD/Control as positive. \textit{Asterisks}$^{*}$: performance significantly different from the binomial majority-class baseline, $p < 0.05$.)}
\label{appendix_fig:LLMs Audio-only 0-Shot}
\end{figure}

\subsection{Kimi-Audio Audio-only Setting}
\vspace{-3mm}

\begin{figure}[H]
\centering
\includegraphics[width=0.7\textwidth]{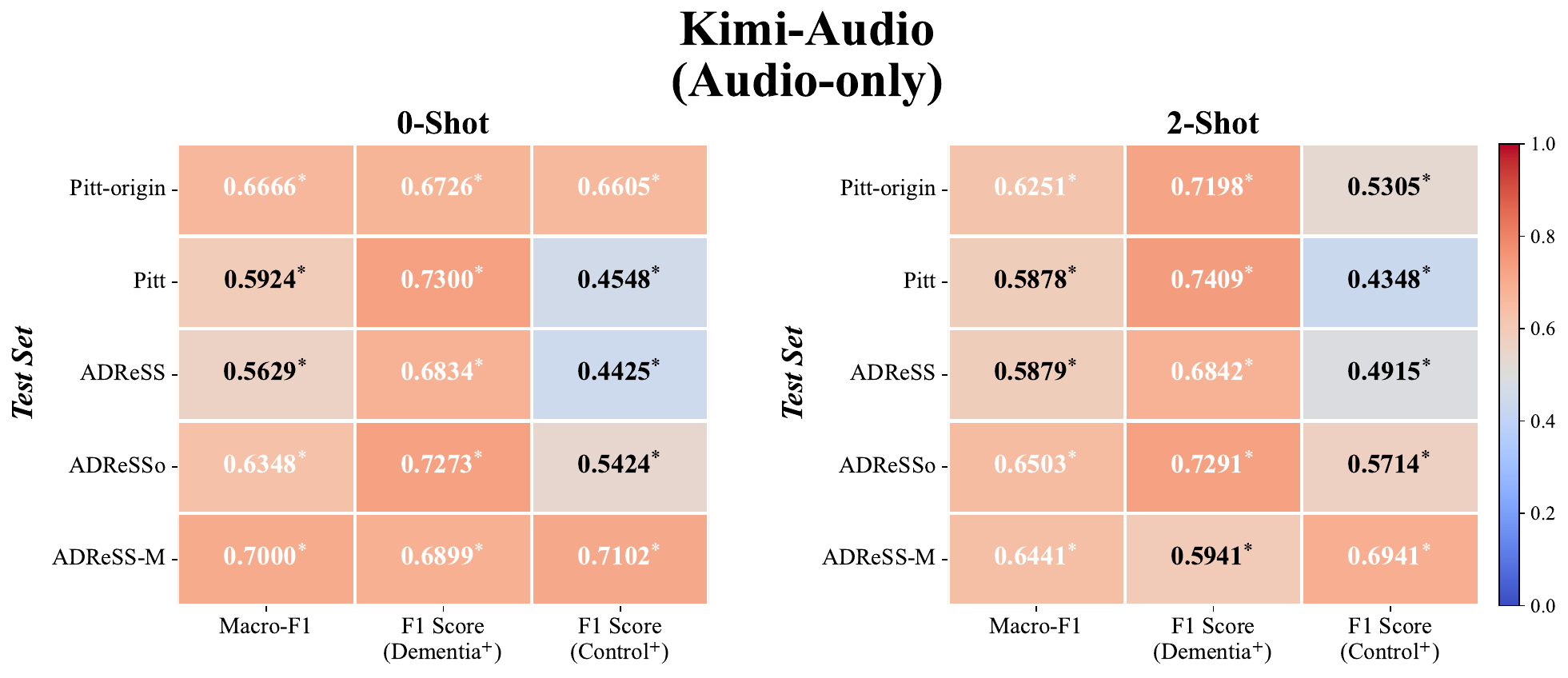}
\vspace{-3mm}
\caption{Kimi-Audio in the audio-only setting. (\textit{AD$^{+}$ / Control$^{+}$}: F1 score with AD/Control as positive. \textit{Asterisks}$^{*}$: performance significantly different from the binomial majority-class baseline, $p < 0.05$.)}
\label{appendix_fig:Kimi-Audio Audio-only}
\end{figure}

\subsection{Kimi-Audio Audio + Transcript Setting}
\vspace{-3mm}

\begin{figure}[H]
\centering
\includegraphics[width=0.7\textwidth]{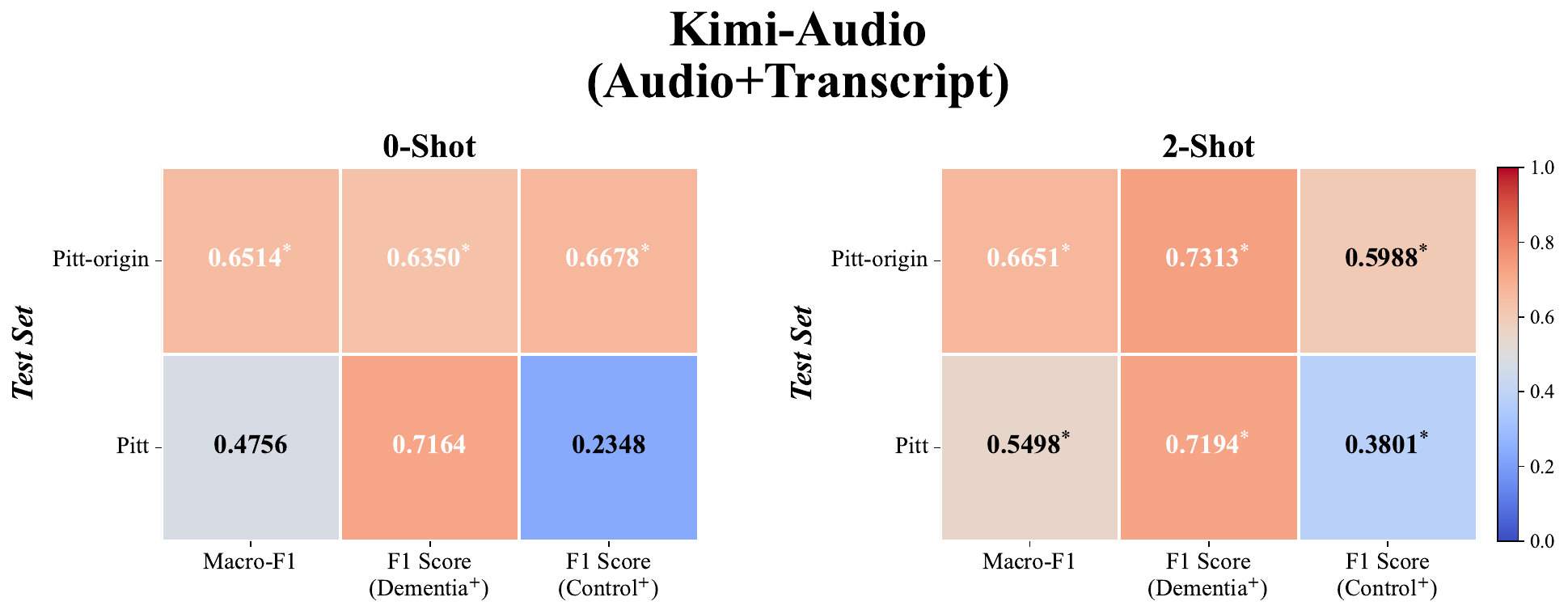}
\vspace{-3mm}
\caption{Kimi-Audio in the audio + transcript setting. (\textit{AD$^{+}$ / Control$^{+}$}: F1 score with AD/Control as positive. \textit{Asterisks}$^{*}$: performance significantly different from the binomial majority-class baseline, $p < 0.05$.)}
\label{appendix_fig:Kimi-Audio Audio+transcript}
\end{figure}

\clearpage
\onecolumn
\section{Model Architectures}
~\label{Appendix: Model Architecture}
\vspace{-4mm}

\begin{figure}[H]
\centering
\includegraphics[width=\textwidth]{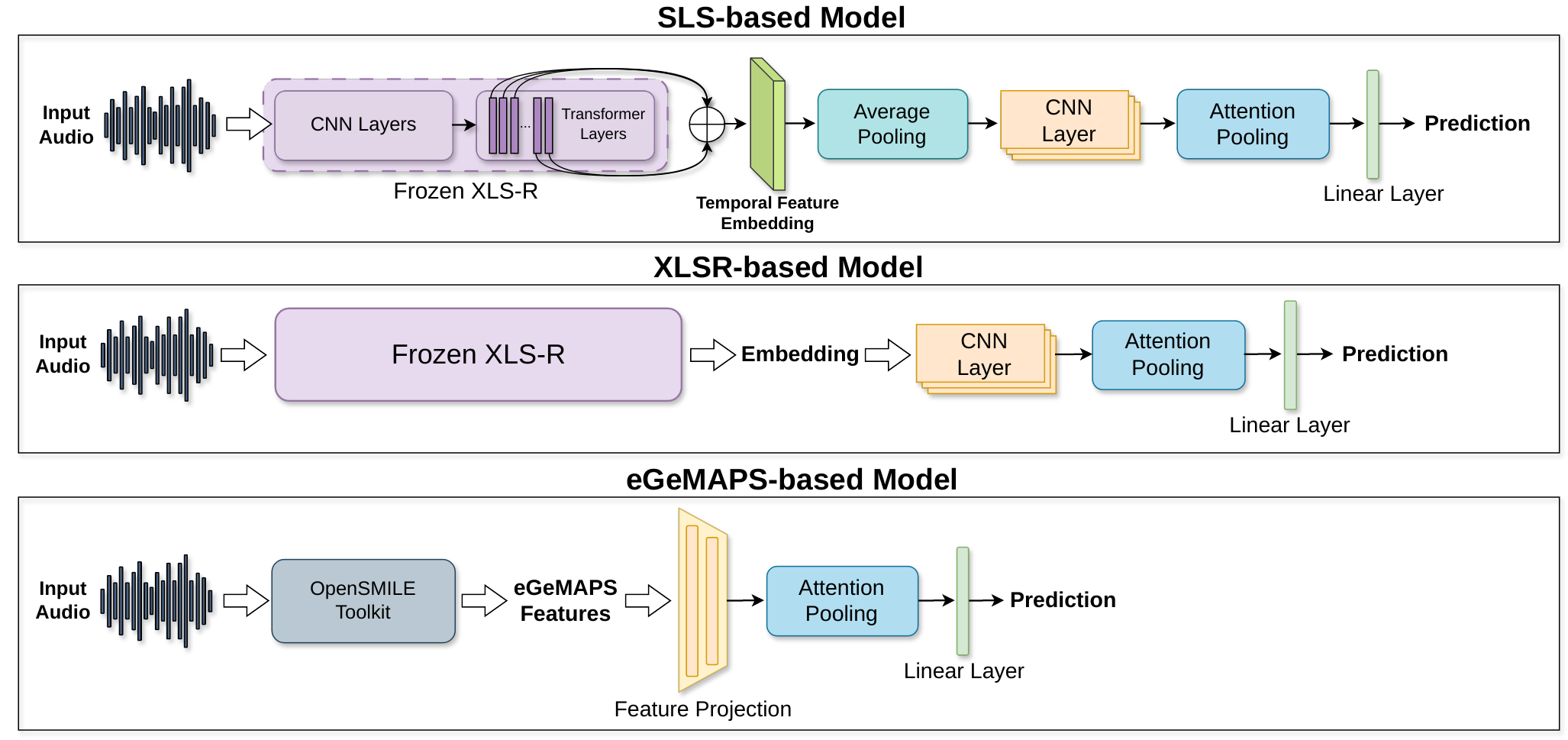}
\vspace{-6mm}
\caption{Model Architectures.}
\label{appendix_fig:model architecture}
\end{figure}

\textbf{SLS-based Model.}
The model takes as input the frame-level representations from all Transformer layers of the parameter-frozen pretrained XLS-R model~\cite{ConneauXLSR2020}, and obtains a weighted fused representation through the Sensitive Layer Selection (SLS)~\cite{zhang2024audio} mechanism. Specifically, the model first applies mean pooling with the padding mask~\cite{VaswaniTransformer2017} along the temporal dimension to the representations of each Transformer layer of XLSR. Subsequently, the weights of different layers are obtained through a linear layer and a sigmoid function. Finally, the model performs a weighted sum of the representations from different layers using the learned layer weights to obtain the weighted fused representation. This representation then sequentially passes through batch normalization, temporal average pooling, a one-dimensional convolutional layer, and attention pooling. Finally, an AD prediction is output through a linear classification layer. 

The SLS-based model contains 169K trainable parameters. All models are trained on NVIDIA RTX 5090 GPUs. Additional implementation details are provided in our released code.

\textbf{XLSR-based Model.}
The audio input is fed into a parameter-frozen pretrained XLSR model~\cite{ConneauXLSR2020} to extract frame-level speech embeddings. The representation then sequentially passes through batch normalization and is aggregated by an attention pooling layer to obtain a temporal feature representation. A padding mask~\cite{VaswaniTransformer2017} is also applied to ensure that padded time steps do not contribute to the pooled representation. Finally, a linear classification layer outputs the AD prediction. 

The XLSR-based model contains 168K trainable parameters. All models are trained on NVIDIA RTX 5090 GPUs. Additional implementation details are provided in our released code.

\textbf{eGeMAPS-based Model.}
The model takes 25-dimensional eGeMAPS~\cite{EybenGeMAPS2015} acoustic features extracted using the OpenSMILE toolkit~\cite{EybenOpenSMILE2010} as input. Each input utterance is first segmented into 10 equal segments, from which 25-dimensional eGeMAPS~\cite{EybenGeMAPS2015} acoustic features are extracted using the OpenSMILE toolkit~\cite{EybenOpenSMILE2010}. Two linear layers are then applied to remap the feature dimensions, followed by an attention pooling layer to aggregate segment-level representations. Finally, a linear classification layer outputs the AD prediction. 

The eGeMAPS-based model contains 6.2K trainable parameters. All models are trained on NVIDIA RTX 5090 GPUs. Additional implementation details are provided in our released code.

\clearpage
\onecolumn
\section{SLS-based Model}
\label{Appendix: SLS-based Model}

\subsection{Generalization Test On Raw Data}
\label{Appendix: SLS Generalization Test On Raw Data}
\vspace{-2mm}

\begin{figure}[H]
\centering
\includegraphics[width=0.5\textwidth]{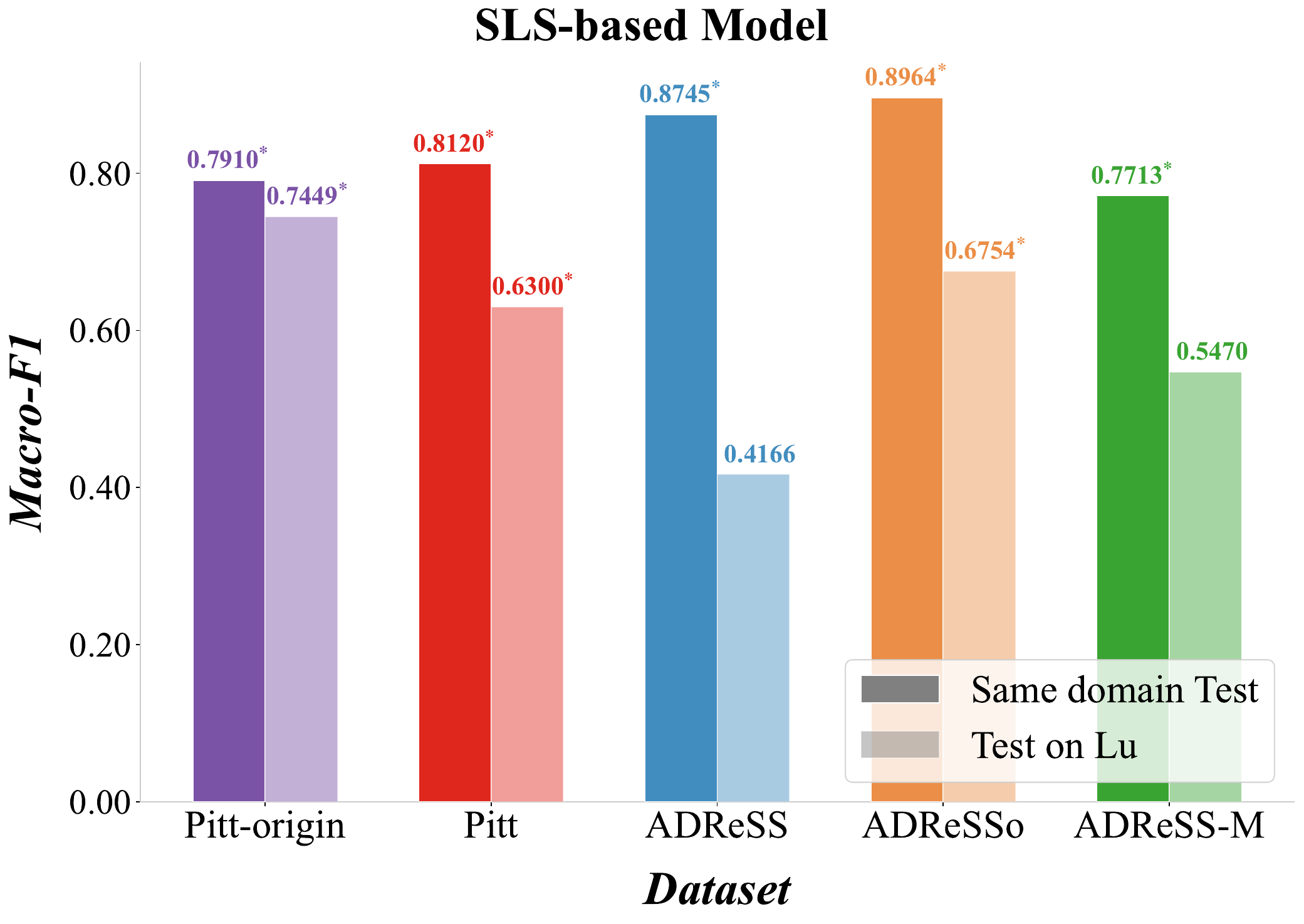}
\caption{SLS-based model generalization test on raw data. (\textit{Same Domain Test}: evaluation on the test split of the same dataset used for training;  \textit{Test on Lu}: evaluation on the Lu dataset. \textit{Asterisks}$^{*}$: performance significantly different from the binomial majority-class baseline, $p < 0.05$.)}
\label{appendix_fig:SLS_unprocessed_generalization}
\end{figure}

\vspace{-2mm}
\subsection{Generalization Test On Speech-enhanced Data}
\vspace{-2mm}

\begin{figure}[H]
\centering
\includegraphics[width=0.7\textwidth]{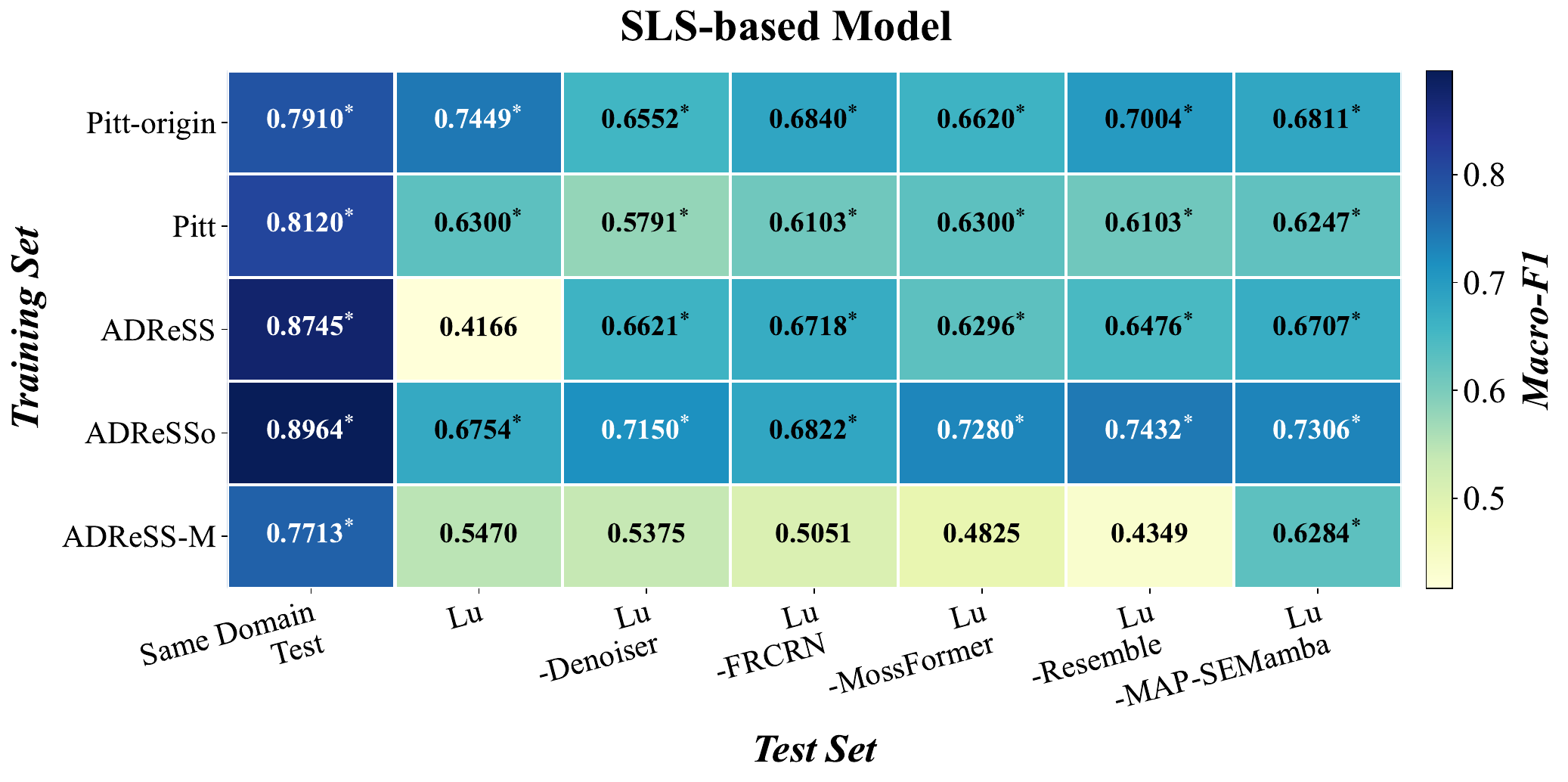}
\caption{SLS-based model generalization test on speech-enhanced data. (\textit{Same Domain Test}: evaluation on the test split of the same dataset used for training. \textit{Asterisks}$^{*}$: performance significantly different from the binomial majority-class baseline, $p < 0.05$.)}
\label{appendix_fig:SLS_generalization}
\end{figure}

\vspace{-4mm}
\subsection{Performance Comparison with Matched Speech Enhancement Methods}
\vspace{-4mm}

\begin{figure}[H]
\centering
\includegraphics[width=0.7\textwidth]{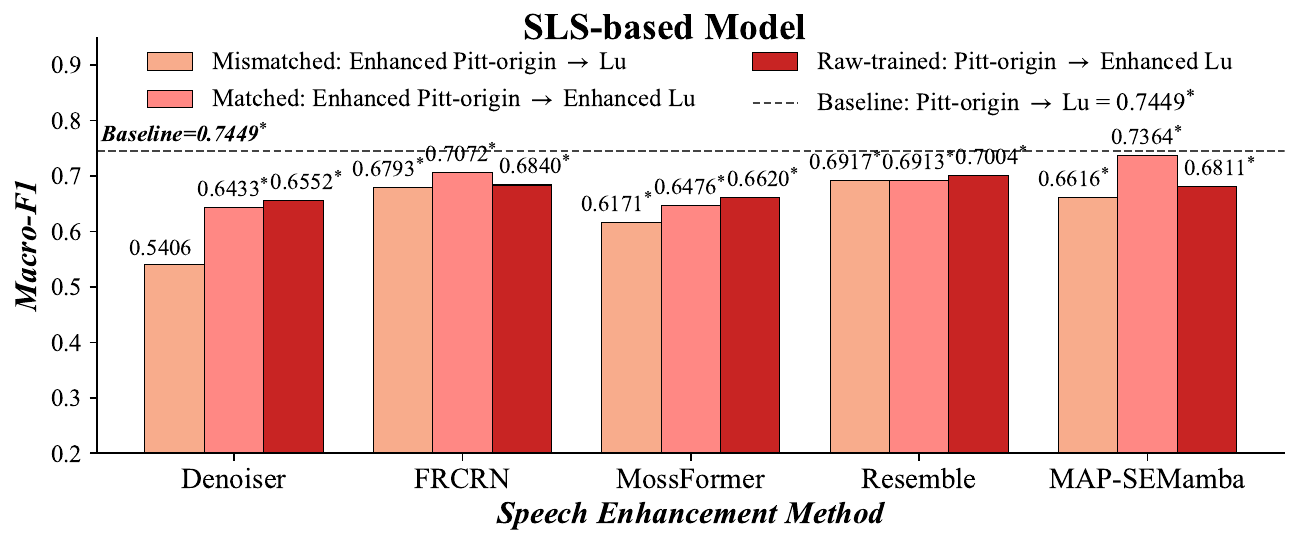}
\caption{SLS-based model performance comparison with matched and mismatched speech enhancement methods. (\textit{Mismatched}: enhanced training set with raw test set; 
\textit{Matched}: the same enhancement for training and test set; 
\textit{Raw-trained}: raw training set with enhanced test set; 
\textit{Baseline}: raw training and raw test set. \textit{Asterisks}$^{*}$: performance significantly different from the binomial majority-class baseline, $p < 0.05$.)}
\label{appendix_fig:SLS_Match}
\end{figure}

\subsection{SLS-based Model All Results}

\begin{figure}[H]
\centering
\includegraphics[width=\textwidth]{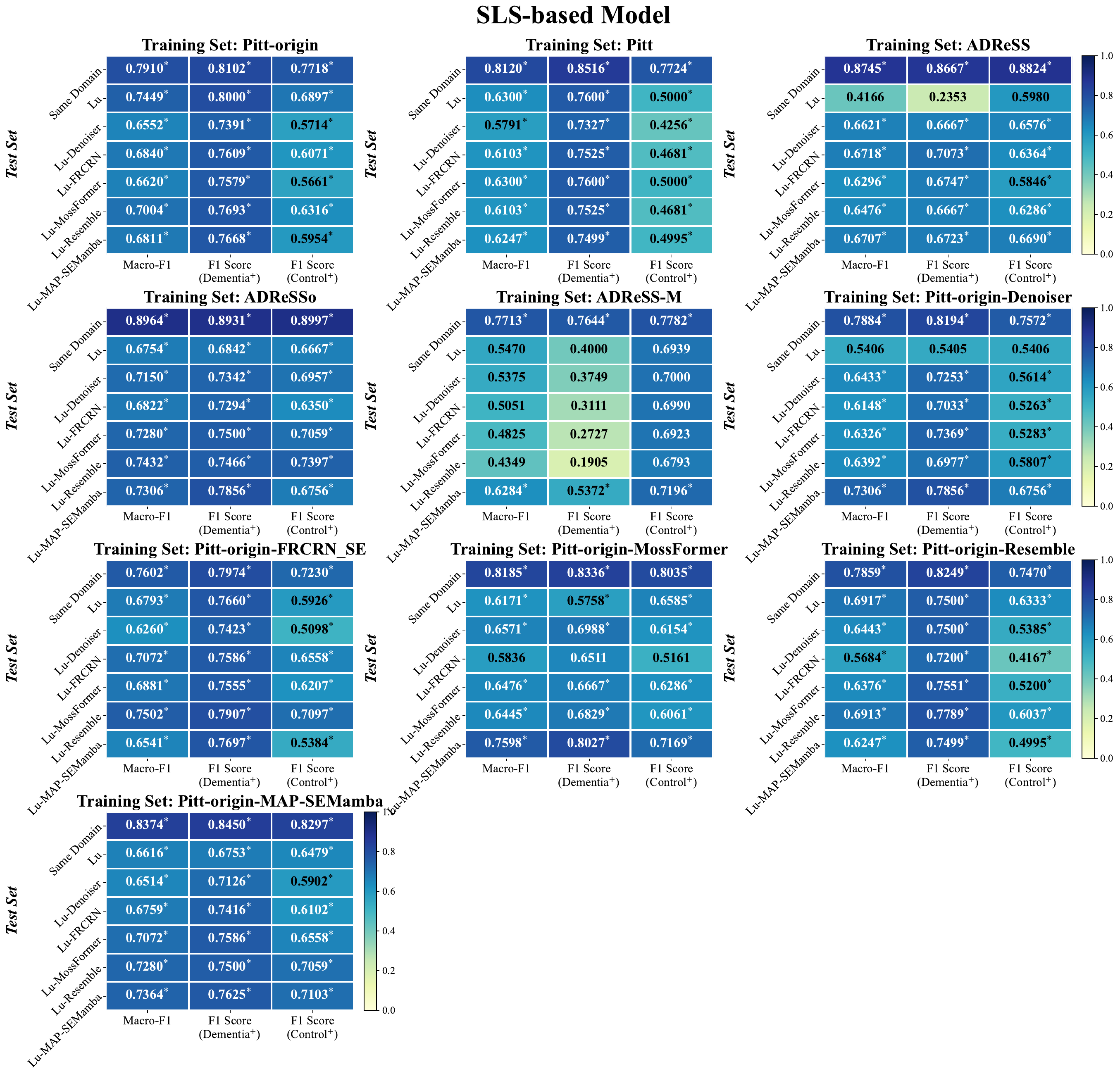}
\vspace{-3mm}
\caption{SLS-based model all results. (\textit{AD$^{+}$ / Control$^{+}$}: F1 score with AD/Control as positive. \textit{Asterisks$^{*}$}: performance significantly different from the binomial majority-class baseline, $p < 0.05$.)}
\label{appendix_fig:SLS-based Model All Results}
\end{figure}

\clearpage
\onecolumn
\section{XLSR-based Model}
\label{Appendix: XLSR-based Model}
\subsection{Generalization Test On Raw Data}
\label{Appendix: XLSR Generalization Test On Raw Data}
\vspace{-2mm}

\begin{figure}[H]
\centering
\includegraphics[width=0.5\textwidth]{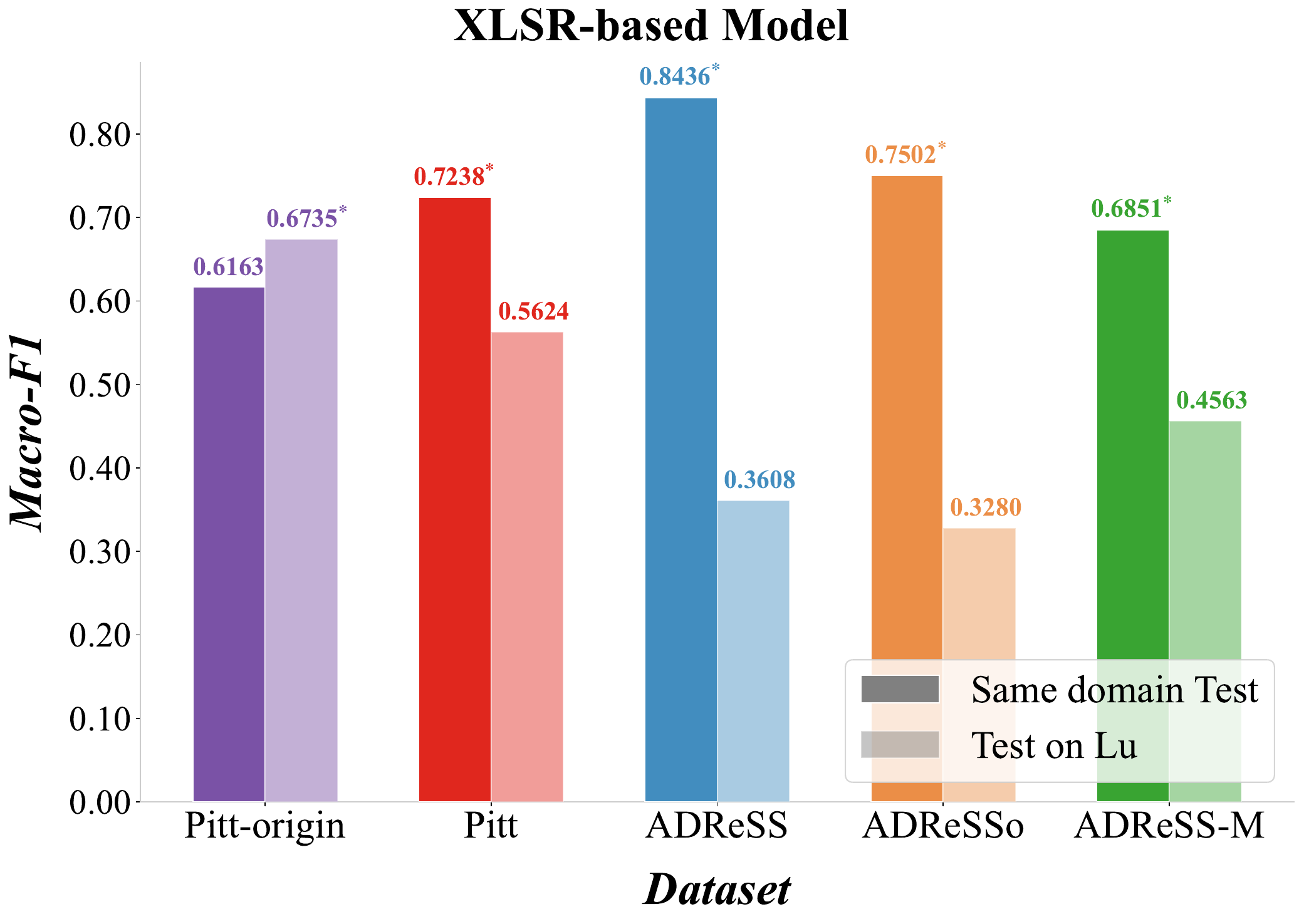}
\caption{XLSR-based model generalization test on raw data. (\textit{Same Domain Test}: evaluation on the test split of the same dataset used for training;  \textit{Test on Lu}: evaluation on the Lu dataset. \textit{Asterisks}$^{*}$: performance significantly different from the binomial majority-class baseline, $p < 0.05$.)}
\label{appendix_fig:XLSR_unprocessed_generalization}
\end{figure}

\vspace{-2mm}
\subsection{Generalization Test On Speech-enhanced Data}
\vspace{-2mm}

\begin{figure}[H]
\centering

\includegraphics[width=0.7\textwidth]{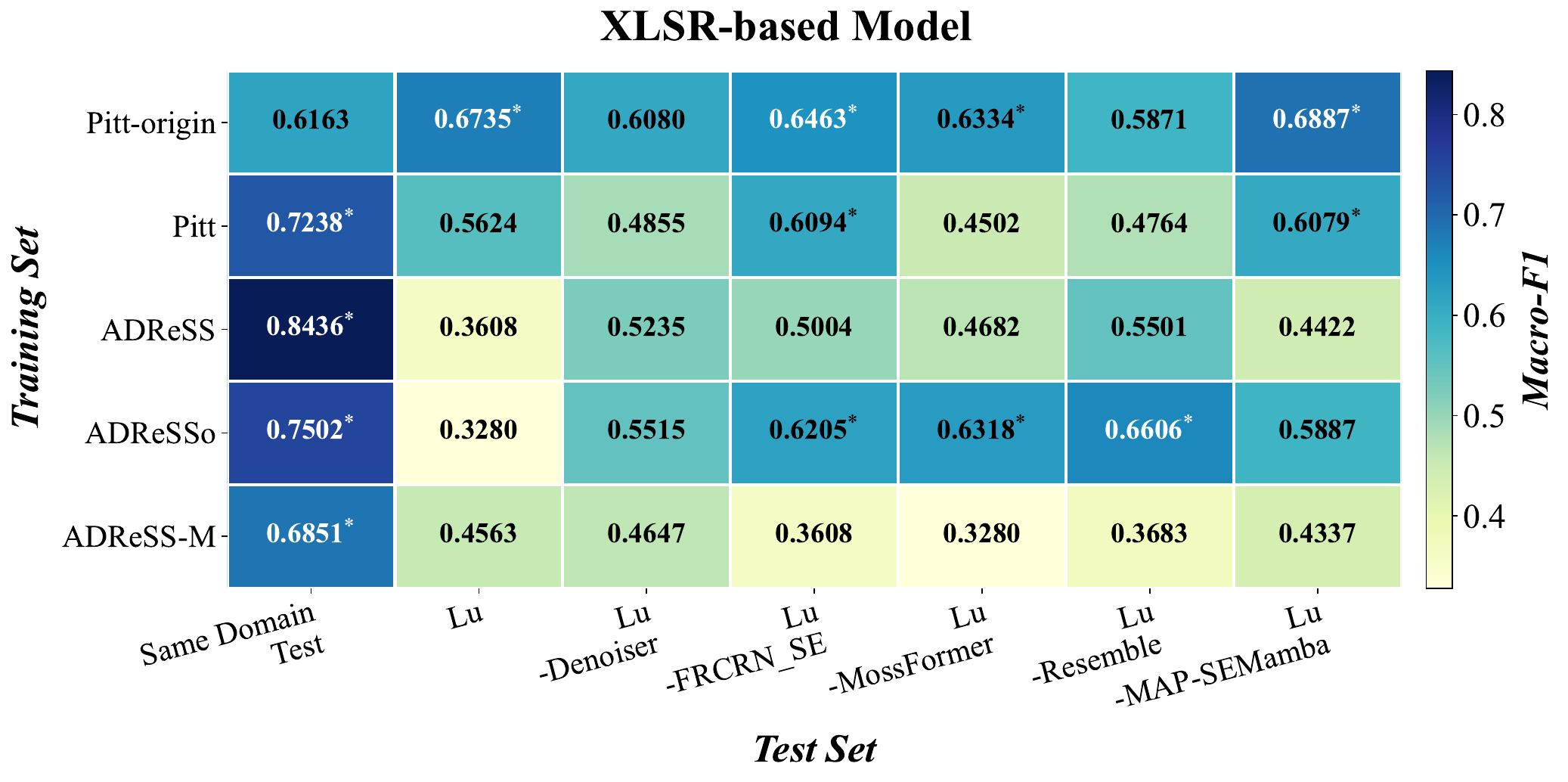}
\caption{XLSR-based model generalization test on speech-enhanced data. (\textit{Same Domain Test}: evaluation on the test split of the same dataset used for training. \textit{Asterisks}$^{*}$: performance significantly different from the binomial majority-class baseline, $p < 0.05$.)}
\label{appendix_fig:xlsr_generalization}
\end{figure}

\vspace{-4mm}
\subsection{Performance Comparison with Matched Speech Enhancement Methods}
\vspace{-4mm}

\begin{figure}[H]
\centering
\includegraphics[width=0.7\textwidth]{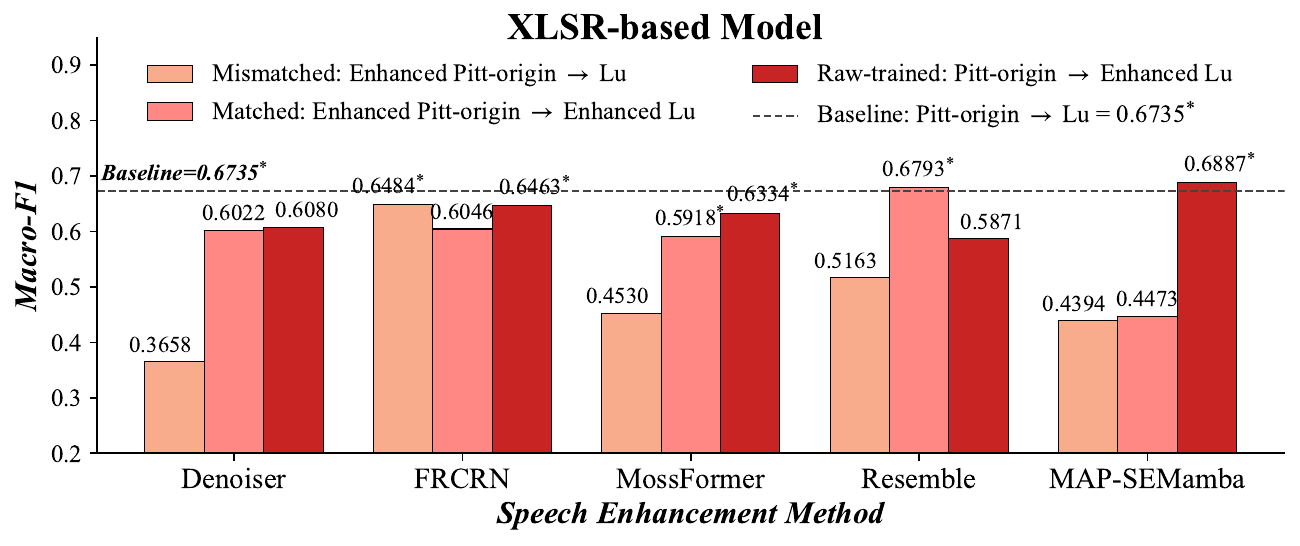}
\caption{XLSR-based model performance comparison with matched and mismatched speech enhancement methods. (\textit{Mismatched}: enhanced training set with raw test set; 
\textit{Matched}: the same enhancement for training and test set; 
\textit{Raw-trained}: raw training set with enhanced test set; 
\textit{Baseline}: raw training and raw test set. \textit{Asterisks}$^{*}$: performance significantly different from the binomial majority-class baseline, $p < 0.05$.)}
\label{appendix_fig:XLSR_Match}
\end{figure}

\subsection{XLSR-based Model All Results}

\begin{figure}[H]
\centering
\includegraphics[width=\textwidth]{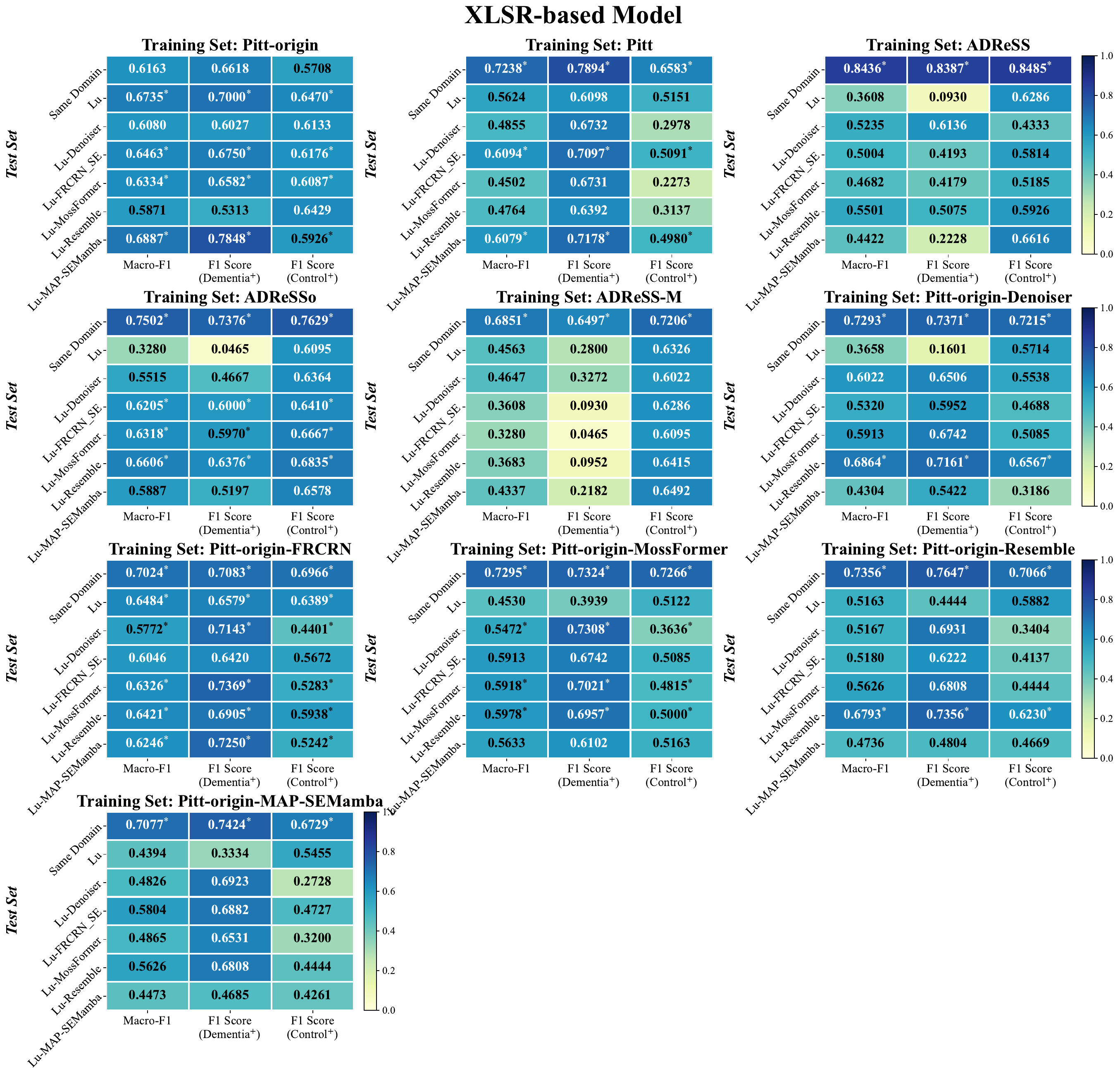}
\vspace{-3mm}
\caption{XLSR-based model all results. (\textit{AD$^{+}$ / Control$^{+}$}: F1 score with AD/Control as positive. \textit{Asterisks$^{*}$}: performance significantly different from the binomial majority-class baseline, $p < 0.05$.)}
\label{appendix_fig:XLSR-based Model All Results}
\end{figure}

\clearpage
\onecolumn
\section{eGeMAPS-based Model}
\label{Appendix: eGeMAPS-based Model}
\subsection{Generalization Test On Raw Data}
\label{Appendix: eGeMAPS Generalization Test On Raw Data}
\vspace{-2mm}

\begin{figure}[H]
\centering
\includegraphics[width=0.5\textwidth]{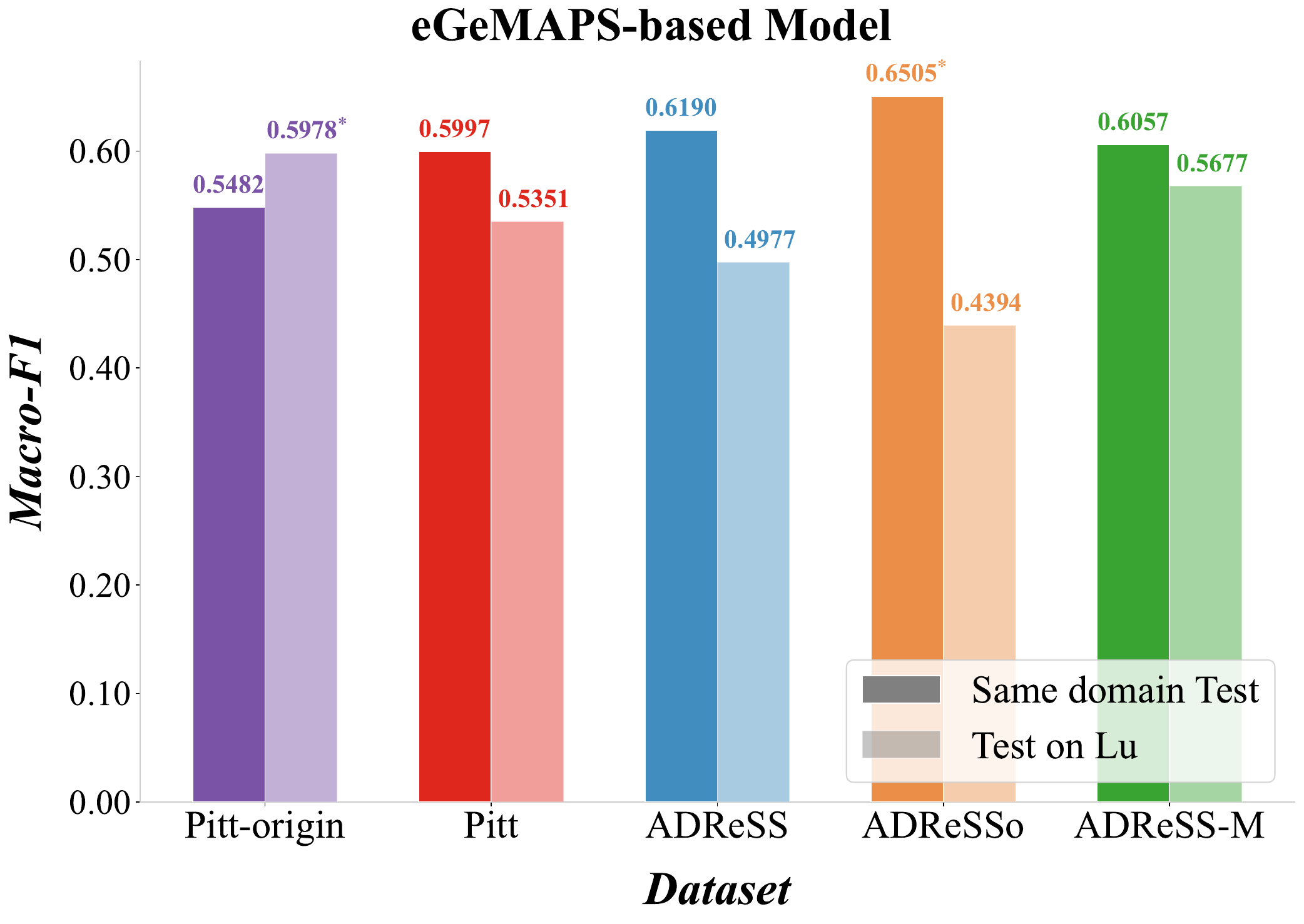}
\caption{eGeMAPS-based model generalization test on raw data. (\textit{Same Domain Test}: evaluation on the test split of the same dataset used for training;  \textit{Test on Lu}: evaluation on the Lu dataset. \textit{Asterisks}$^{*}$: performance significantly different from the binomial majority-class baseline, $p < 0.05$.)}
\label{appendix_fig:egemaps_unprocessed_generalization}
\end{figure}

\vspace{-2mm}
\subsection{Generalization Test On Speech-enhanced Data}
\vspace{-2mm}

\begin{figure}[H]
\centering
\includegraphics[width=0.7\textwidth]{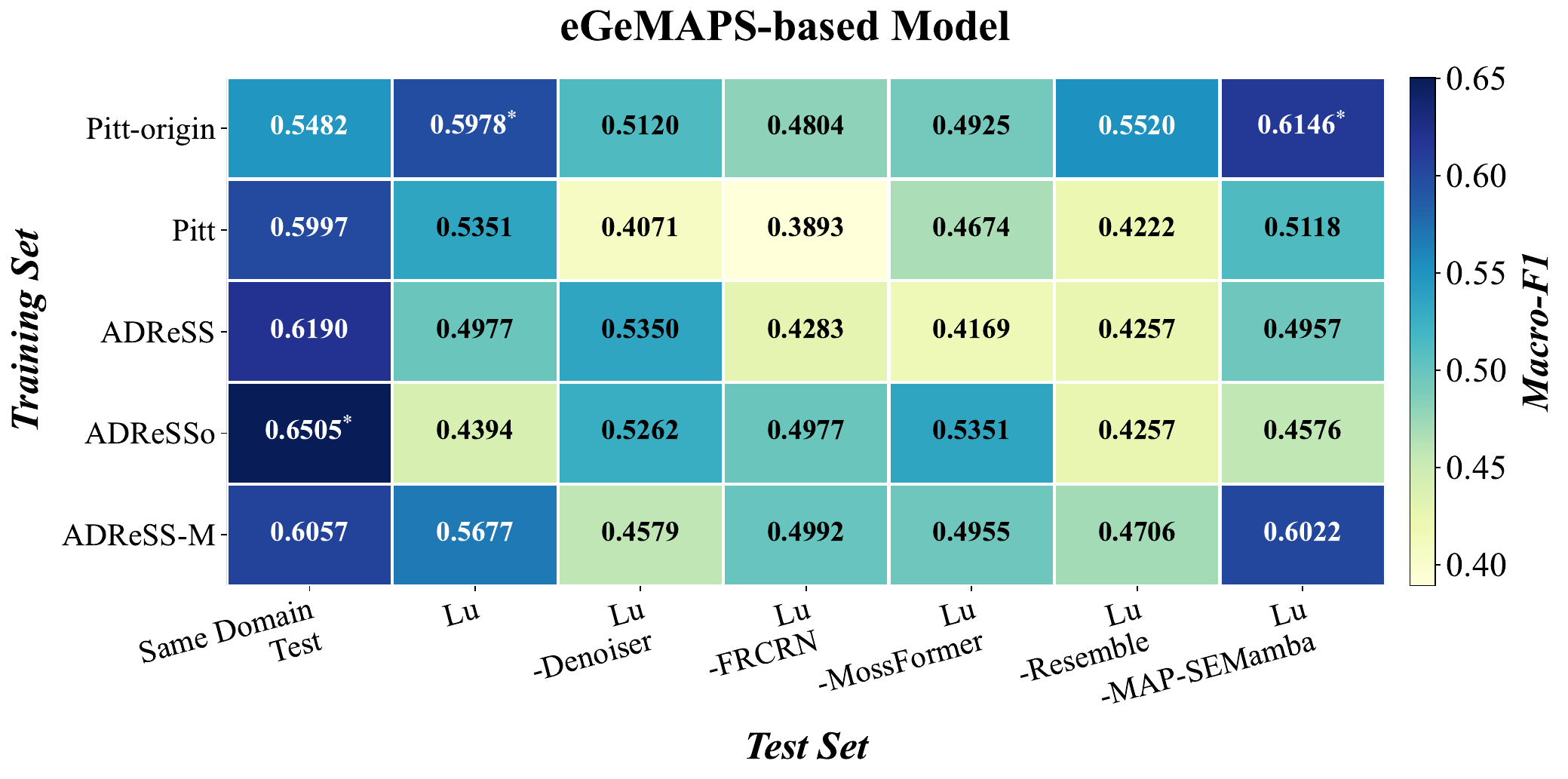}
\caption{eGeMAPS-based model generalization test on speech-enhanced data. (\textit{Same Domain Test}: evaluation on the test split of the same dataset used for training. \textit{Asterisks}$^{*}$: performance significantly different from the binomial majority-class baseline, $p < 0.05$.)}
\label{appendix_fig:egemaps_generalization}
\end{figure}

\vspace{-4mm}
\subsection{Performance Comparison with Matched Speech Enhancement Methods}
\vspace{-4mm}

\begin{figure}[H]
\centering
\includegraphics[width=0.7\textwidth]{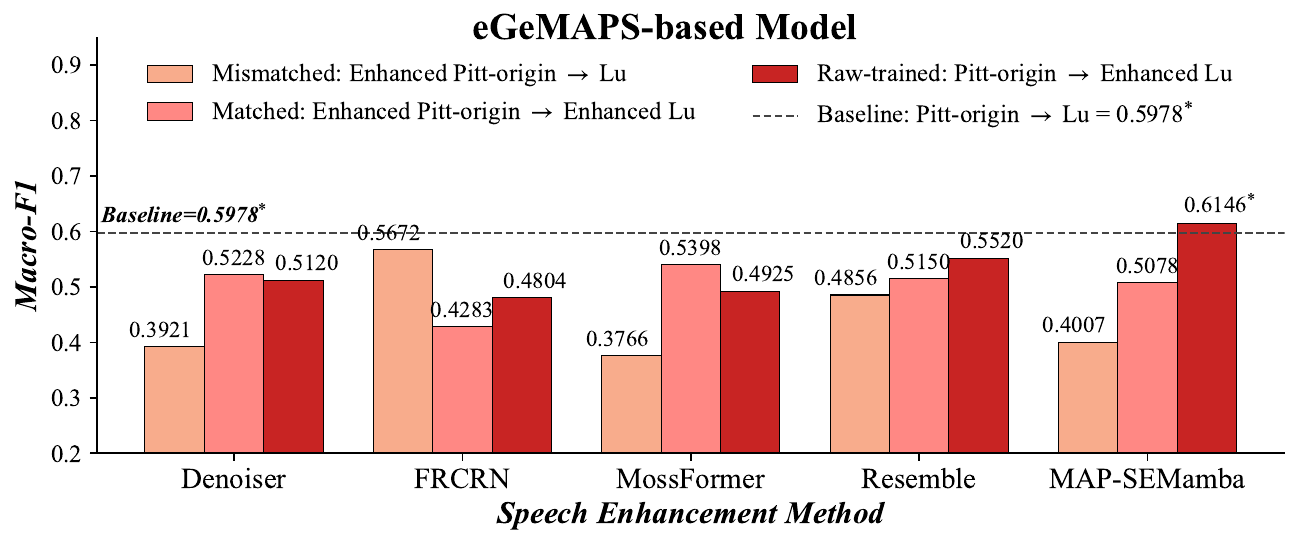}
\caption{eGeMAPS-based model performance comparison with matched and mismatched speech enhancement methods. (\textit{Mismatched}: enhanced training set with raw test set; 
\textit{Matched}: the same enhancement for training and test set; 
\textit{Raw-trained}: raw training set with enhanced test set; 
\textit{Baseline}: raw training and raw test set. \textit{Asterisks}$^{*}$: performance significantly different from the binomial majority-class baseline, $p < 0.05$.)}
\label{appendix_fig:egemaps_Match}
\end{figure}

\subsection{eGeMAPS-based Model All Results}

\begin{figure}[H]
\centering
\includegraphics[width=\textwidth]{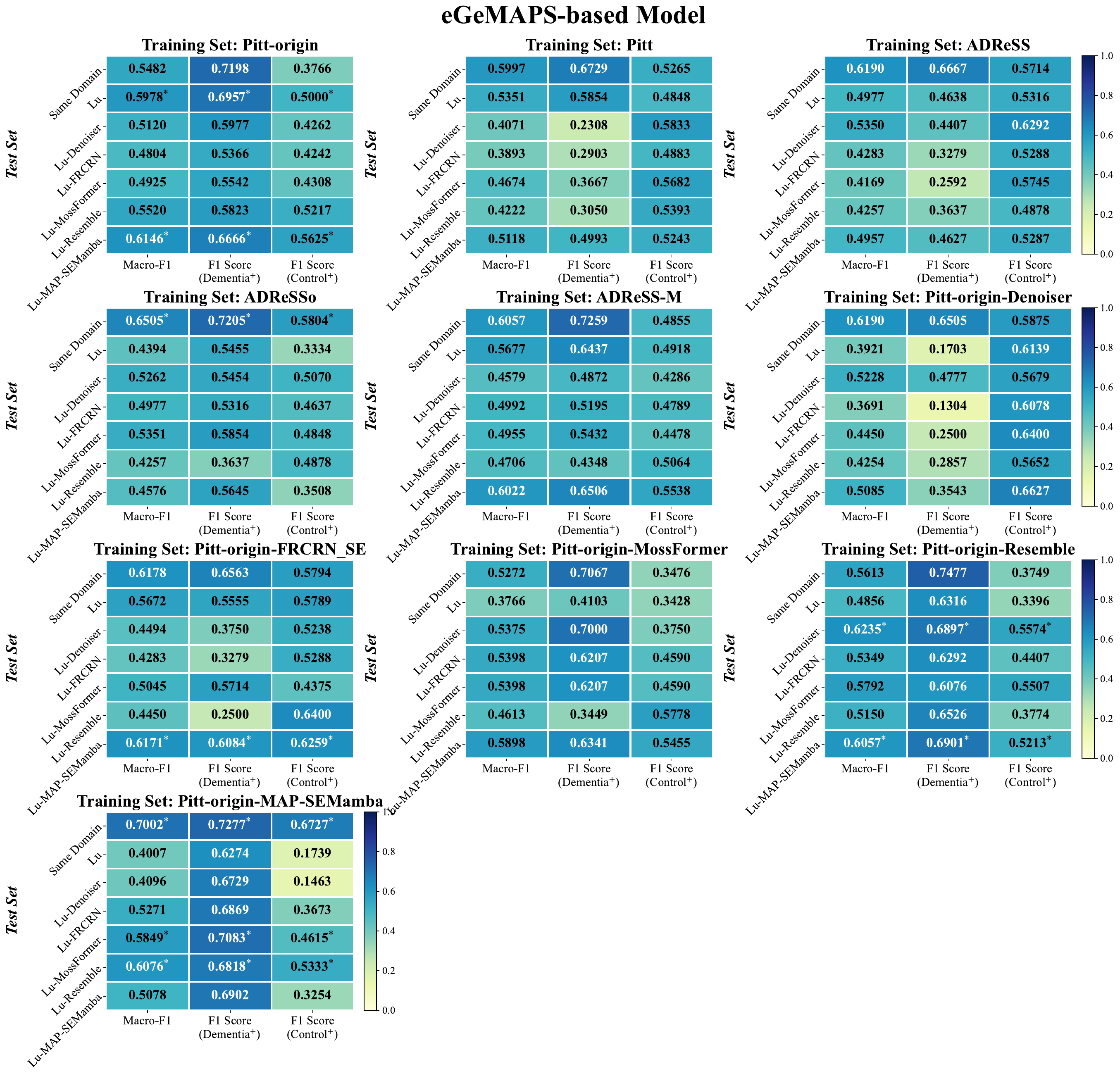}
\vspace{-3mm}
\caption{eGeMAPS-based model all results. (\textit{AD$^{+}$ / Control$^{+}$}: F1 score with AD/Control as positive. \textit{Asterisks$^{*}$}: performance significantly different from the binomial majority-class baseline, $p < 0.05$.)}
\label{appendix_fig:eGeMAPS-based Model All Results}
\end{figure}

\end{document}